\documentclass[pra,twocolumn,aps,showpacs,superscriptaddress,floatfix,showpacs]{revtex4-1}

\usepackage{graphicx}
\usepackage{dcolumn}
\usepackage{bm}
\usepackage{latexsym}
\usepackage{amssymb}
\usepackage{amsmath}
\usepackage{braket}

\newcommand{\opsandwich}[3]{\left< #1|#2|#3 \right >}

\newcommand{\expec}[1]{\left < #1\right >}

\newcommand{\lp}{\left ( }
\newcommand{\rp}{\right ) }
\newcommand{\lb}{\left [ }
\newcommand{\rb}{\right ] }

\newcommand{\beq}{\begin{eqnarray*}}
\newcommand{\eeq}{\end{eqnarray*}}

\newcommand{\be}{\begin{eqnarray}}
\newcommand{\ee}{\end{eqnarray}}

\usepackage[usenames,dvipsnames]{color}

\usepackage[normalem]{ulem}

\usepackage[colorlinks,bookmarks=false,citecolor=blue,linkcolor=red,urlcolor=blue]{hyperref}

\begin{document}

\title{  SU$(N)$ magnetism in chains of ultracold alkaline-earth-metal atoms: Mott transitions and quantum correlations }

\date{\today}

\author{Salvatore R. Manmana}
\affiliation{JILA, (University of Colorado and NIST), and Department of Physics, University of Colorado, Boulder, Colorado 80309-0440, USA}
\author{Kaden R. A. Hazzard}
\affiliation{JILA, (University of Colorado and NIST), and Department of Physics, University of Colorado, Boulder, Colorado 80309-0440, USA}
\author{Gang Chen}
\affiliation{JILA, (University of Colorado and NIST), and Department of Physics, University of Colorado, Boulder, Colorado 80309-0440, USA}
\author{Adrian E. Feiguin}
\affiliation{Department of Physics and Astronomy, University of Wyoming, Laramie, WY 82071, USA}
\author{Ana Maria Rey}
\affiliation{JILA, (University of Colorado and NIST), and Department of Physics, University of Colorado, Boulder, Colorado 80309-0440, USA}

\begin{abstract}
We investigate  one dimensional  SU$(N)$ Hubbard chains at zero temperature, which can be emulated with ultracold alkaline earth atoms, by using the density matrix renormalization group (DMRG), Bethe ansatz (BA), and bosonization. 
We compute experimental observables and use the DMRG  to benchmark the  accuracy of the Bethe ansatz for $N>2$ where the BA is only approximate. 
In the worst case, we find a relative error $\varepsilon \lesssim 4\%$ in the  BA  ground state energy  for $N \leq 4$ at filling $1/N$, which is due to the fact that BA improperly treats the triply and higher occupied states.
Using the DMRG for $N \leq 4$ and the BA for large $N$, we determine the regimes of validity of strong- and weak-coupling perturbation theory for all values of $N$ and in particular,  the parameter range in which the system is well described by a SU$(N)$ Heisenberg model at filling $1/N$. 
We find this depends only weakly on $N$.
We investigate  the Berezinskii-Kosterlitz-Thouless phase transition  from a Luttinger liquid to a Mott-insulator by computing  the fidelity susceptibility and the Luttinger parameter $K_\rho$ at $1/N$ filling.  
The numerical findings give strong evidence that  the fidelity susceptibility develops a minimum  at a critical interaction strength  which is found  to occur at  a finite positive value  for $N>2$.
\end{abstract}

\maketitle

\section{Introduction}
\label{sec:intro}
The  SU$(N)$ Hubbard model  describes $N$-flavor fermions hopping on a lattice with  flavor-independent onsite interactions.
The model is a generalization of  the conventional  SU$(2)$ Hubbard model  introduced in the 1960s for the theoretical description of itinerant ferromagnets in spin $1/2$ systems \cite{Hubbard_orig,Hubbard_orig_Gutzwiller,Hubbard_orig_Kanamori}. 
It has attracted considerable theoretical attention in recent years \cite{Affleck:1988p1082,Assaraf:1999p728,Wu:2003p1524,Assaraf:2004p1546,Honerkamp:2004p1077,Lecheminant:2005p1550,
Szirmai:2005p683,Assaad:2005p1538,Wu:2005p1588,Szirmai:2006p682,Rapp:2007p1574,Zhao:2007p1611,Buchta:2007p713,
Lecheminant:2008p1553,Rapp:2008p1572,Szirmai:2008p712,Capponi:2008p1533,Molina:2009p1525,
Gorelik:2009p1066,Roux:2009p1597,Azaria:2009p1616,Inaba:2009p6398,Miyatake:2010p6400,Klingschat:2010p1566,Ulbricht:2010p1544,Szirmai:2010p1569,Nonne:2010p1517,Hazzard:2010p1516,Inaba:2011p6402,Inaba:2010p6401}. 
In the strong interacting regime  and at $1/N$ filling (one particle per site), the low energy physics of the SU$(N)$ Hubbard model  is captured by an effective SU$(N)$ Heisenberg model in which the charge degrees of freedom is frozen and only the spin degrees of freedom are  allowed to fluctuate. 
There is  a long history of studies of   SU$(N)$  spin systems \cite{Sutherland:1975p1590,Johannesson:1986p6403,Johannesson:1986p6404,Read:1989p2014,Affleck:1988p1082}. 
Their initial motivation was to better understand the usual SU$(2)$  antiferromagnets since SU$(N)$ spins are analytically tractable in the large-$N$ limit.  
These studies have found rich phase diagrams \cite{Sutherland:1975p1590,Affleck:1991p1627,PhysRevB.61.12112,Harada:2003p1619,Li:2004p1632,Paramekanti:2007p1604,Fuhringer:2008p1609,Kolezhuk:2008p1614,
Aguado:2009p1534,Kawashima:2007p1607,Thomale:2007p1577,Xu:2008p1625,Rachel:2009p762,Rachel:2009p763,Beach:2009p1599,Tu:2011p6399}, exhibiting antiferromagnetic ordering, valence-bond solids and in 1D exotic spin-nematic phases \cite{spin_nematics1,spin_nematics2,Li:2007p1276,Toth:2010p1006,Lauchli:2006p1030,Tsunetsugu:2006p1070,PhysRevB.74.144426,PhysRevB.76.220404,PhysRevB.83.184433} and generalizations of the AKLT state \cite{Affleck:1987p1591,Affleck:1988p1615,Read:1989p2014,Greiter:2007p1573,
Greiter:2007p1527,Rachel:2008p1539,Hermele:2009p1034,Rachel:2009p1633,Alet:2011p1582,Lou:2009p1562}, among others.  
However, as no exact SU$(N)$ models have existed in nature, these predictions were considered as a theoretical playground.

The recent discovery that the SU$(N)$ Hubbard model  describes ultracold gases of alkaline earth atoms on optical lattices~\cite{Gorshkov:2010p1052,
Taie:2010p2009,Cazalilla:2009p1000,Fukuhara:2009p1589} brings these considerations into a new perspective and has spurred theoretical and experimental interest.
Having in mind this specific experimental implementation, exotic new phases  -- such as chiral spin liquids \cite{PhysRevLett.59.2095,PhysRevB.39.11413} -- have been predicted~\cite{Hermele:2009p1034} which can be of relevance for the realization of topological quantum computers \cite {Kitaev2006,Nayak2008}.

A first step towards the experimental observation of the rich spin  physics  in ultracold alkaline earth atoms  is the precise knowledge of the parameter regime in which the SU$(N)$ Heisenberg model describes the low energy physics of the  full SU$(N)$ Hubbard model. In this paper, we address this question. By combining numerical density matrix renormalization group (DMRG) \cite{White:1992p2171,White:1993p2161,dmrgbook,Schollwock:2005p2117} simulations with Bethe ansatz (BA), we are able to predict the onset of the validity of the spin-models for all values of $N$. We pursue three main purposes with this paper.
Firstly, we address the mainly theoretical aspects of the validity of the approximate BA for $N>2$.  
Secondly, using quantum information measures we determine the value of the critical interaction $U_c$ at which the Mott transition \cite{gebhard} takes place.
Thirdly, we use these insights to provide predictions of  the minimal value of $U_s$ at which the atoms behave as spin systems. Those predictions are relevant for ongoing experiments with ultracold alkaline earth atoms. 
We  also discuss numerical results for the on-site occupancies, the various correlation functions,  momentum distributions, and their structure factors which can be accessed in current experiments.

We present these results as follows.
In Sec.~\ref{sec:modelsandobservables}, we introduce the SU$(N)$ Hubbard model and the SU$(N)$  Heisenberg model  and discuss their  realization in ultracold  alkaline earth atoms on optical lattices. We also describe the methods we use to study these models, and their connection to experimentally relevant observables.
We discuss in Sec.~\ref{sec:BAtreatment} the BA treatment of the SU$(N)$ Hubbard chains for $N\geq2$, and in Sec.~\ref{sec:fidelity} the properties of the fidelity susceptibility $\chi(U)$ in the vicinity of a quantum critical point.
In Sec.~\ref{sec:compareDMRGBA}, we compare the BA and DMRG results for the ground state energy per site in the thermodynamic limit and discuss the relative error of the BA results as a function of $U$.
In Sec.~\ref{sec:fidsusc} we address the Mott transition using our results for the fidelity susceptibility.
Based on the results of these sections, we turn to experimentally relevant aspects in Sec.~\ref{sec:largeU}.
In particular, we identify $U_s \gtrsim 12t$ for all values of $N$ by comparing energies and spin correlation functions of the Hubbard and Heisenberg systems in Secs.~\ref{sec:HubbardHeisenbergEnergies} and~\ref{sec:correlresults}. 
In Secs.~\ref{sec:correlresults} and \ref{sec:structurefactors}, we present our DMRG results for the on-site occupancies, and the correlation functions and their structure factors and discuss possibilities to access these quantities in experiments.
Section~\ref{sec:summary} summarizes our findings.

\section{Models, Experimental Realization, Methods and Observables}
\label{sec:modelsandobservables}

\subsection{Models}
\label{sec:models}
We treat the 1D fermionic SU$(N)$ Hubbard model described by the Hamiltonian

\begin{equation}
\mathcal{H} = - t \sum\limits_{\langle ij \rangle, \, \alpha}  (f^\dagger_{\alpha,i} f^{\phantom{\dagger}}_{\alpha,j} +h.c.) + \frac{U}{2} \sum\limits_{i, \alpha \neq \alpha^\prime} f^\dagger_{\alpha,i} f^\dagger_{\alpha^\prime,i} f^{\phantom{\dagger}}_{\alpha^\prime,i} f^{\phantom{\dagger}}_{\alpha,i} \label{hub}
\end{equation}
using Bethe ansatz and  DMRG.
The fermionic operator $f_{\alpha,i}$ ($f^{\dagger}_{\alpha,i}$) destroys (creates) a particle of flavor $\alpha$ on lattice site $i$, with $\alpha = 1,\ldots,N$, and $\sum_{\langle i, j \rangle}$ denotes a sum over nearest neighbor sites $i$ and $j$. 
In the limit of large $U/t$, the effective model obtained by second order degenerate perturbation theory is the SU$(N)$ Heisenberg model
\begin{equation}
\mathcal{H} = \frac{2 t^2}{U} \sum_{\langle ij \rangle, \alpha\beta} S^\beta_\alpha(i) \, S^\alpha_\beta(j),
\label{eq:SUNHeisenberg}
\end{equation}
with the spin operators $S_\alpha^\beta(i) = f^\dagger_{\alpha,i} f^{\phantom{\dagger}}_{\beta,i}$ which
satisfy the SU$(N)$ algebra $[S_\alpha^\beta(i),S_\gamma^\delta(j)] = \delta_{ij} (\delta_{\beta \gamma}
S_\alpha^\delta - \delta_{\alpha \delta} S_\beta^\gamma)$ and hence form the generators of the SU$(N)$ symmetry.
We consider the deep Mott insulator (MI) state, in which the system satisfies the local constraint of having one particle per site,
$\sum_\alpha f^\dagger_{\alpha,i} f^{\phantom{\dagger}}_{\alpha,i} = 1$,
so that the spin on each site transforms in the fundamental representation of SU$(N)$.
Thus the number of sites
needed to form a singlet is $N$, leading to rich and exotic behavior \cite{Read:1989p2014,Read1989,PhysRevLett.81.3527}.
In Ref.~\onlinecite{Hermele:2009p1034} it was shown that on a square lattice, besides the expected antiferromagnetic phase, there are valence bond solids, and -- most interestingly -- topologically ordered chiral spin liquids.

\subsection{Experimental Realization in Systems of Alkaline Earth Atoms}

Alkaline earth atoms belong to the second column of the periodic table and have two-outer valence electrons.
These and other atoms with similar atomic structure such as Yb  have unique atomic properties  that make them attractive candidates for new types of atomic clocks \cite{Ludlow:2008p1642,Lemonde:2009p1671,Lemke:2009p2007}, quantum simulation \cite{Gorshkov:2010p1052,
Cazalilla:2009p1000} and quantum information applications \cite{Daley:2008p1523,Hayes:2007p2003,Gorshkov:2009p1620}. In  their ground state, $^1S_0$, the electronic degrees
of freedom have neither spin nor orbital angular momentum ($J=0$) and the nuclear spin ($I$) is thus decoupled from the electronic state. Only fermionic isotopes have $I>0$ \cite{nuclear_physics} and these are the focus of our study. 
This decoupling not only allows one to independently manipulate nuclear and electronic  degrees of freedom, but also implies that the ${}^1S_0$  s-wave scattering lengths are independent of the nuclear spin. 
The nuclear-spin-dependent variation of the scattering lengths is expected to be  smaller than $\sim 10^{-9}$ \cite{Gorshkov:2010p1052}. 
This leads to the SU($N$) symmetric models treated in this paper, with $N \leq 2 I +1$.

A fundamental consequence of the SU$(N)$ symmetry  is the conservation of the total number of atoms with nuclear spin projection  $\alpha$. 
This means that an atom with large I, e.g. ${}^{87}$Sr $(I = 9/2)$, can reproduce the dynamics of atoms with lower $I$ if one takes an initial state with no population in the extra levels. 
This feature of SU$(N)$ symmetry is in stark contrast to the case of weaker SU$(2)$ symmetry exhibited in alkali atoms, where the dependence of scattering lengths on the total spin of the two colliding particles  allows for spin changing collisions, due to the finite hyperfine interactions. 

The many-body Hamiltonian that  describes  cold fermionic alkaline-earth atoms in the  $^1S_0$ state loaded in the lowest band  of an optical lattice is Eq.~\eqref{hub} with  $t = - \int d^3 \mathbf{r} w (\mathbf{r}) (-
\frac{\hbar^2}{2 M} \nabla^2 + V(\mathbf{r}))
w(\mathbf{r}-\mathbf{r_0})$,
where $\mathbf{r_0}$ is the separation of two nearest neighbors, and $U = \frac{4 \pi \hbar^2}{M}a_{gg} \int
d^3\mathbf{r}w^4(\mathbf{r})$, with $M$ the mass, $a_{gg}$ the nuclear spin independent $^1S_0$  scattering length and $w (\mathbf{r})$ the Wannier functions of an atom in a lattice potential  $ V(\mathbf{r})$ \cite{Gorshkov:2010p1052}.  
In order to allow motion only along one direction the lattice confinement along the other two must be strong to suppress tunneling in the course of the experiment. 

\subsection{Approximate Bethe Ansatz for $N>2$}
\label{sec:BAtreatment}
The Bethe ansatz  exactly solves many models, for example the SU$(2)$ Heisenberg model, SU$(N)$ continuum fermions, and the SU$(N)$ Heisenberg model \cite{Takahashi_BA,Sutherland:1975p1590}.  
However, the SU$(N)$ Hubbard model is not solvable with standard Bethe ansatz techniques: since lattice sites may be occupied by more than two particles, it is impossible to reduce the many-particle problem to two-particle scattering events.
Haldane and Choy developed a ``natural" generalization of the SU$(2)$ Hubbard model Bethe ansatz equations to SU$(N)$ symmetry, given in Refs.~\onlinecite{choy:some_1980,haldane:sun-bethe_1980}.
The obtained solution, however only approximates the Hubbard model, as discussed below. 
Appendix~\ref{sec:appendixA} presents the details of the calculation.

As shown there, we obtain for the ground state energy per site
\be
E_{\text{BA}} &=& -2t\int_{-k_0}^{k_0} \! dk\, \cos(k) \rho_c(k) \label{eq:BA-energy},
\ee
with the pseudomomenta $k$ and charge rapidity distribution $\rho_c(k)$, where $k_0$ is determined by the density via $$n=\int_{-k_0}^{k_0}\! dk\, \rho_c(k). $$
In this paper we have restricted to the balanced case with equal population of each of the $N$ spin components, where the Bethe ansatz equations simplify to~\cite{lee:anomalous_1989}
\be
\rho_c(k) &=& \frac{1}{2\pi} + \cos(k) \int_{-k_0}^{k_0} \! dk' \, \rho_c(k') G_N(\sin(k)-\sin(k')), \nonumber \\
G_N(x) &=& \frac{1}{2\pi} \int\! d\omega\, e^{-i\omega x} e^{-U|\omega|/4} \frac{\sinh\lb\lp N-1\rp U\omega/4 \rb }{\sinh \lp N U\omega/4\rp}\label{eq:simplified-BA-balanced}
\ee
Even though we work with a balanced gas, for our purposes it is equally convenient to simply solve the equations Eq.~\eqref{eq:BA-densities-SUN-T0} numerically, as described in Appendix~\ref{sec:appendixA}.


While there have been numerous theoretical studies of this approximate Bethe ansatz~\cite{choy:some_1980,haldane:sun-bethe_1980,choy:failure_1982,lee:anomalous_1989,schlottmann:spin_1991,schlottmann:metal-insulator_1992,frahm:critical_1993}, no precise quantification of its accuracy was available.
In Sec.~\ref{sec:compareDMRGBA} we provide such a quantification by comparing to numerical DMRG results.

Haldane and Choy have argued that the Schr{\"o}dinger equation is exactly satisfied only for configurations where less than three particles occupy a site~\cite{choy:failure_1982}. This is the reason the BA is approximate.

The simplest quantification of the approximate nature of the SU$(N)$ Bethe ansatz solution is found by considering the  SU$(N)$ Hubbard model's three particle problem.
In particular~\cite{choy:failure_1982}, one finds that
\be
\opsandwich{\mathbf{x}}{H-E(\mathbf{k})}{\mathbf{v}{k}} &=& \frac{U^2}{4t} f(\mathbf{k})\opsandwich{\mathbf{x}}{{\hat P}_3}{\mathbf{k}},\label{eq:3ptcl-BA}
\ee
with 
\[
\begin{split}
f(\mathbf{k}) &= \\
&\left[ \cos\left(\frac{k_1+k_2}{2}\right) \cos\left( \frac{k_2+k_3}{2}\right) \cos\left( \frac{k_3+k_1}{2}\right) \right]^{-1}, 
\end{split}
\]
and defining the energy $E(\mathbf{k}) =$ $ -2t \lb  \cos(k_1) \right.$ $  + \cos(k_2)$ $ \left. + \cos(k_3)\rb$, the three-particle per site projection operator ${\hat P}_3\equiv \delta_{x_1 x_2}\delta_{x_2 x_3}$, the state $\ket{\mathbf{k}}$ to be the three particle Bethe ansatz wavefunction characterized by pseudomomenta $k_1$, $k_2$, and $k_3$, and $\ket{\mathbf{x}}$ the state with particles at positions $x_1$, $x_2$, and $x_3$.
It is illustrative here, however, to note that even when $U=0$, and thus the number of configurations with triple or larger occupancies is large, the approximate Bethe ansatz is nevertheless exact due to the $U^2/t$ prefactor.

Additionally the Bethe ansatz is exact when no configurations have site occupancies with greater than two particles per site.
This includes the $1/N$ filling $U=\infty$ (hard core) limit where it reproduces the exact BA solution of the SU$(N)$ Heisenberg model,  and the dilute
limit $\expec{n}\ll 1$ where it reproduces the behavior of SU$(N)$ $\delta$-function interacting particles.

\subsection{Details of the DMRG calculation}

Due to the large on-site Hilbert spaces, the DMRG is restricted to $N \leq 5$.
In addition, for the gapless systems treated in the following, a large number of density-matrix eigenstates $m$ must be kept.
We therefore treat systems only up to $L=216$ lattice sites and keep up to $m=4000$ density matrix eigenstates. We discuss results obtained with open boundary conditions (OBC) since the DMRG is most efficient in this case.

Note that an additional restriction appears when computing the fidelity susceptibility $\chi(U)$ [Eq.~(\ref{eq:chi})] discussed below.
The fidelity $\mathcal F(U)$ [Eq.~(\ref{eq:fidelity})] is very close to one, so that $1-\mathcal F(U) \sim 10^{-5}$.
It is therefore necessary to achieve the corresponding convergence in the energy and a discarded weight which is significantly smaller than this number.
These requirements allow to compute $\chi(U)$ reliably only for systems with up to $L=192$ sites for $N=2$ and $L=48$ sites for $N=3$, keeping up to $m=4000$ and performing 10 sweeps.
Additional results for larger system sizes and for $N=4$ show the same qualitative features, but we will not use them for the finite size extrapolations since the convergence of these calculations does not match the requirements for a reliable analysis of $\chi(U)$.
Note that the use of non-abelian quantum numbers \cite{ian4,ian3,ian2,ian1} might help in future studies to realize larger system sizes and treat larger values of $N$.

\subsection{Correlation functions and their structure factors}
\label{sec:correlfunctions}

In this section we introduce the correlation functions which we are going to discuss in more detail in Sec.~\ref{sec:correlresults}.
The goals are to compare the spin correlation functions of the Heisenberg and the Hubbard systems  when varying $U$ in order to identify the Heisenberg regime of the Hubbard chains, and to provide the structure factors of spin and charge correlation functions since these are accessible to experiments as discussed below.

For the Heisenberg model, we compute directly
\begin{equation}
S^{\rm H}(l,m)_{\alpha,\beta} = \left\langle S^\beta_\alpha(l) \, S^\alpha_\beta(m) \right\rangle.
\label{eq:spincorrsheisenberg}
\end{equation}
Note that due to the SU$(N)$ symmetry the correlation functions along the spin quantization axis and perpendicular to it are identical, so that it is sufficient to consider only Eq.~(\ref{eq:spincorrsheisenberg}).
The situation would, however, be different in the presence of symmetry-breaking external fields or population imbalance.

For the Hubbard chains, by  expanding the spin operators $S_\alpha^\beta$ in terms of the Fermi operators $f_{\alpha,i}$, and taking into account the SU$(N)$ symmetry, one finds that the spin correlation functions and the associated structure factor can be obtained from the difference of density correlation functions of two identical and two different flavors of the particles,
\begin{equation}
\begin{split}
&S(l,m) = \left\langle n_l^\alpha n_m^\alpha \right\rangle - \left\langle n_l^\alpha n_m^\beta \right\rangle \\
&\mathcal S(k) = \frac{1}{L} \sum\limits_{l,m} S(l,m) \, e^{i k (l-m)},
\end{split}
\label{eq:spincorrs}
\end{equation}
where $n_l^\alpha = f_{\alpha,l}^{\dagger} f_{\alpha,l}^{\phantom{\dagger}}$. 
Note that here summation over repeated indices is not implied.  

Complementary to this, density correlation functions and their structure factors are 
\begin{equation}
\begin{split}
&N_{\alpha,\alpha} (l,m) =  \left\langle n_l^\alpha n_m^\alpha \right\rangle - \left\langle n_l^\alpha\right\rangle \left\langle n_m^\alpha \right\rangle\\
&\mathcal{N}_{\alpha,\alpha} (k) = \frac{1}{L} \sum\limits_{l,m} N_{\alpha,\alpha}(l,m) \, e^{i k (l-m)},
\end{split}
\label{eq:alphaalphacorrs}
\end{equation}
and
\begin{equation}
\begin{split}
&N_{\alpha,\beta} (l,m) =  \left\langle n_l^\alpha n_m^\beta \right\rangle - \left\langle n_l^\alpha\right\rangle \left\langle n_m^\beta \right\rangle\\
&\mathcal{N}_{\alpha,\beta} (k) = \frac{1}{L} \sum\limits_{l,m} N_{\alpha,\beta}(l,m) \, e^{i k (l-m)},
\end{split}
\label{eq:alphabetacorrs}
\end{equation}
for the correlations between particles of the same and of two different species, respectively.
Again, due to the SU$(N)$ symmetry, it is sufficient to restrict to two observables: one with $\alpha = \beta$, and one with $\alpha \neq \beta$ (these are otherwise independent of $\alpha$ and $\beta$).
In addition, it is useful to introduce the correlation function and the structure factor of the total density,
\begin{equation}
\begin{split}
&N(l,m) = \left\langle N^{\rm total}_l N^{\rm total}_m \right\rangle
- \left\langle N^{\rm total}_l \right\rangle \left\langle N^{\rm total}_m \right\rangle \\
&\mathcal{N} (k) = \frac{1}{L} \sum\limits_{l,m} N(l,m) \, e^{i k (l-m)},
\end{split}
\label{eq:denscorrs}
\end{equation}
with $ N^{\rm total}_i = \sum_\alpha n_i^\alpha$.
Further information is provided by the one-particle density matrix (OPDM) and the momentum distribution function,
\begin{equation}
\begin{split}
&\varrho_{\alpha,\alpha}(l,m) = \left\langle f_{\alpha,l}^{\dagger} f_{\alpha,m}^{\phantom{\dagger}} \right\rangle\\
&n_\alpha(k) = \frac{1}{L} \sum\limits_{l,m} \varrho_{\alpha,\alpha}(l,m) e^{i k (l-m)}.
\end{split}
\label{eq:OPDM}
\end{equation}
Note that due to the SU$(N)$ symmetry $\varrho_{\alpha,\beta}(i,j) = 0$ for $\alpha \neq \beta$ so that it is sufficient to consider only $\varrho_{\alpha,\alpha}(l,m)$.

While for the experiments on optical lattices it is possible to measure the correlation functions in real space using {\it in situ} techniques \cite{Gemelke:2009p1898,Bakr:2009p2641,Bakr:2010p1984,Sherson:2010p2701,Trotzky:2010p930,Weitenberg:2011p2530}, the structure factors in momentum space are  easier to access.
In particular, the momentum distribution function $n_\alpha(k)$ can be accessed  via time of flight measurements \cite{Bloch:2005p988,Bloch:2008p943}, the spin structure factor deep in the Mott insulator phase $\mathcal S(k)$ is accessible via noise correlations in the time of flight measurements \cite{PhysRevA.78.012330}, 
and the density structure factor $\mathcal N(k)$ can be measured using Bragg scattering \cite{Bloch:2008p943,Veeravalli:2008p2984,PhysRevA.72.023407}. By applying a magnetic field, the nuclear spin states can be spectroscopically distinguished in transitions to electronic excited states.
 The reason is that the Lande $g$-factor of the excited state significantly
differs from that of the ground state ({\em e.g.}\/ $\sim 60\%$ for
strontium, Sr \cite{Boyd2007}) and in a biased magnetic field, the
various Zeeman transitions  have different resonant frequencies,
 as demonstrated in Ref.~\onlinecite{Boyd2006}.
 Hence Bragg scattering with light with frequency near a resonance for state $\alpha$   measures $\mathcal N_{\alpha, \alpha}(k)$.  In the balanced SU$(N)$ gas, from the knowledge of $\mathcal N(k)$ and  $\mathcal N_{\alpha, \alpha}(k)$ one can compute $\mathcal N_{\alpha,\beta}$.  In the more general -- possibly spin imbalanced -- case (not considered in the present manuscript), one can directly measure $N_{\alpha\beta}$.  To accomplish this, one uses probe light with a frequency $\omega$ where more than one spin species has response.  Tuning $\omega$ tunes the relative response of the different spin flavors, and thus tunes the correlations that are measured by the probe, in a well-characterized way.  Ref.~\onlinecite{corcovilos_detecting_2010} analyzes the  SU$(2)$ case in detail.

\subsection{Fidelity and Fidelity Susceptibility in the small $U$ limit}
\label{sec:fidelity}

From Bethe ansatz it is well known that for $N=2$ at $1/N$ filling there is
a Berezinskii-Kosterlitz-Thouless  transition in the charge sector from a gapless metallic (Luttinger liquid) phase to a gapped Mott-insulating phase at $U=0$ \cite{Hubbard_Book,gebhard,solyom_book}.
While there are strong indications that the transition happens at a value $U_c > 0$ for $N>2$ \cite{Assaraf:1999p728},
Ref.~\onlinecite{Buchta:2007p713} questioned this based on quantum information measures computed from DMRG,
and a scenario in which $U_c$ is either zero or very close to zero was proposed.
Here, we investigate the behavior of the system at small values of $U$ by computing the fidelity
which we define as the overlap between ground states at neighboring points of the coupling constants (here the on-site interaction $U$) \cite{PhysRevE.74.031123}
\begin{equation}
 \mathcal{F}(U) = \left|\langle\psi_0(U)| \psi_0(U+dU)\rangle\right|
\label{eq:fidelity}
\end{equation}
and the fidelity susceptibility
\begin{equation}
\chi(U) =  \frac{2\big[1 - \mathcal{F}(U)\big]}{L \, dU^2},
\label{eq:chi}
\end{equation}
also known as the fidelity metric \cite{PhysRevLett.99.100603}.
For many phase transitions $\chi$ is expected to diverge in the thermodynamic limit (TL), and it has been shown that it possesses a clear signature of the critical point already for rather small systems \cite{PhysRevA.77.012311,PhysRevB.77.245109,PhysRevB.76.180403}.
However, in Ref.~\onlinecite{PhysRevLett.99.095701} the singular part of the fidelity metric in the vicinity of quantum critical points was analyzed by a general scaling argument, and it was found that the singular part of the fidelity susceptibility does not necessarily diverge at a critical point.
Instead, it can have a minimum at the critical point.
For the SU$(N)$ Hubbard model at $1/N$ filling, there is spin-charge separation and each sector is described by a Luttinger liquid theory. 
For two independent theories the fidelity factorizes and the fidelity susceptibility is additive  \cite{PhysRevB.76.180403}, which leads to the relation 
\begin{equation}
\chi (U)= \chi_{\rho} (U) + \chi_{\sigma} (U)
\end{equation}
holds.
Since the spin sector realizes a Luttinger liquid for all values of $U$, we can safely reproduce the analysis of Refs.~\onlinecite{PhysRevB.76.180403,PhysRevB.78.115410} which leads to the relation
\begin{equation}
\chi_{\sigma} = \frac{1}{8} \left(\frac{d\, \log\big(K_{\sigma}(U)\big)}{dU}\right)^2 .
\label{eq:chi_Krho}
\end{equation}
Here, $K_\sigma(U)$ is the spin Luttinger parameter, and due to the SU$(N)$ symmetry, $K_\sigma(U) \equiv 1$ for all values of $U\geq0$ and $N$.
In the Luttinger liquid region, Eq.~(\ref{eq:chi_Krho}) holds also for the relation between $\chi_\rho(U)$ and $K_\rho(U)$.
$\chi(U)$ computed via Eq.~(\ref{eq:chi}) reveals the behavior of the charge sector and can be applied to investigate the Mott transition.
As discussed in more detail in Sec.~\ref{sec:fidsusc}, we find numerically that $\chi(U)$ is minimized at the phase transition.

\section{Comparison between Bethe ansatz and DMRG}
\label{sec:compareDMRGBA}

\begin{figure}[t]
\includegraphics[width=0.5\textwidth]{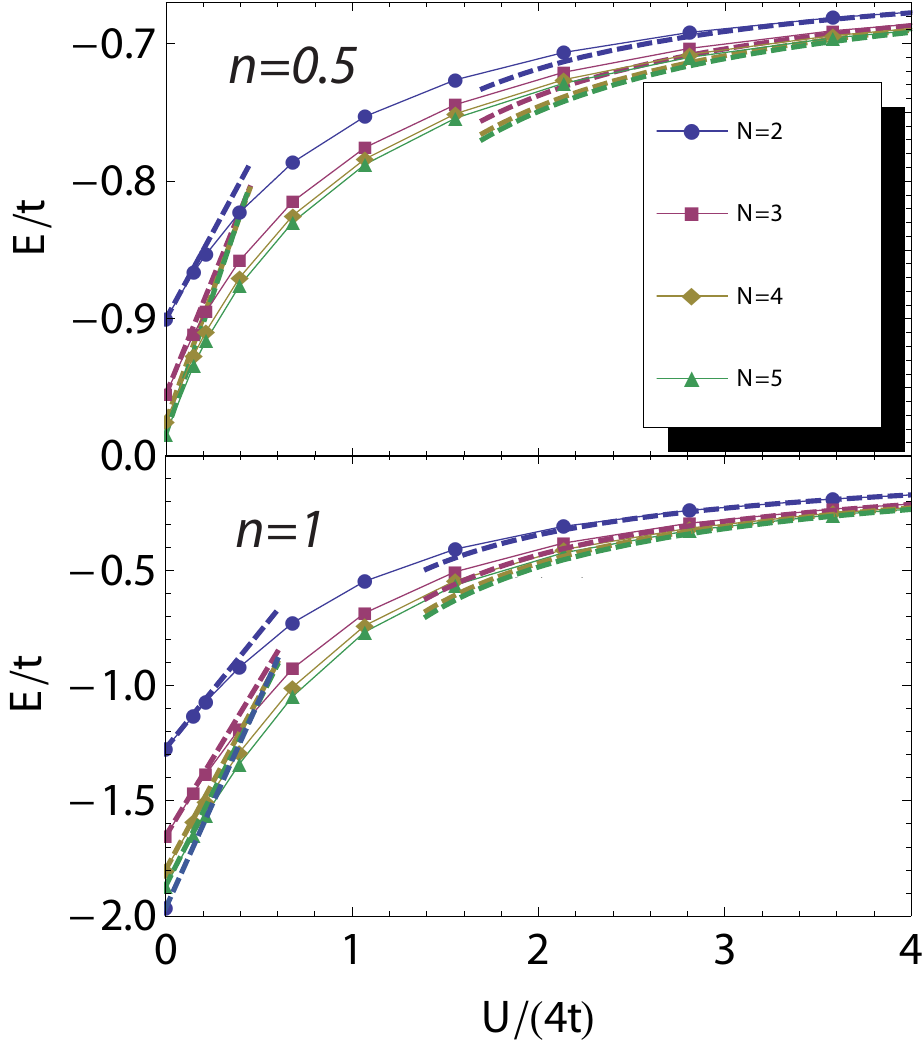}
\caption{(Color online) Bethe ansatz results for the energies in the thermodynamic limit for $N = 2, \, 3, \, 4,\, 5$ (from top to bottom) for density $n=1/2$  ($1/(2N)$ filling, top) and $n=1$ ($1/N$ filling, bottom). The dashed lines indicate the weak coupling and the strong coupling limits.}
\label{fig:BAenergies}
\end{figure}

Figure~\ref{fig:BAenergies} shows the BA results for the ground state energy per site in the thermodynamic limit up to $N=10$ for one atom per site $n=1$ ($1/N$ filling) and for half an atom per site $n=0.5$ ($1/(2N)$ filling). 
Due to the expensive numerics, we consider DMRG results only for $N \leq 4$ and the same values of the filling.
Both DMRG and the BA show the same qualitative behavior for all $N$:
for large values of $U$, the energy asymptotically approaches a constant, while for small values of $U$ it is proportional to $U$.
This suggests the presence of two different regimes and a crossover region or phase transition between them.
We come back to this point in Sec.~\ref{sec:largeU} where we discuss in more detail the parameter regime in which the systems behave as SU$(N)$ Heisenberg spin chains.
Note that at $1/N$ filling, $n=1$, upon increasing $N$, the energies quickly approach an asymptote, so that the curves for $N=3$ and $N=4$ in Fig.~\ref{fig:BAenergies} are basically indistinguishable.  This shows the particles become effectively distinguishable quickly for density $n=1$.
As discussed in Ref.~\onlinecite{lee:anomalous_1989}, in the limit $N \to \infty$ a generalization of the solution of the Lieb-Liniger equation is obtained, so that the SU$(N)$ Hubbard chain in this limit can be regarded as a generalized continuum boson system.

\begin{figure}[t]
\includegraphics[width=0.5\textwidth]{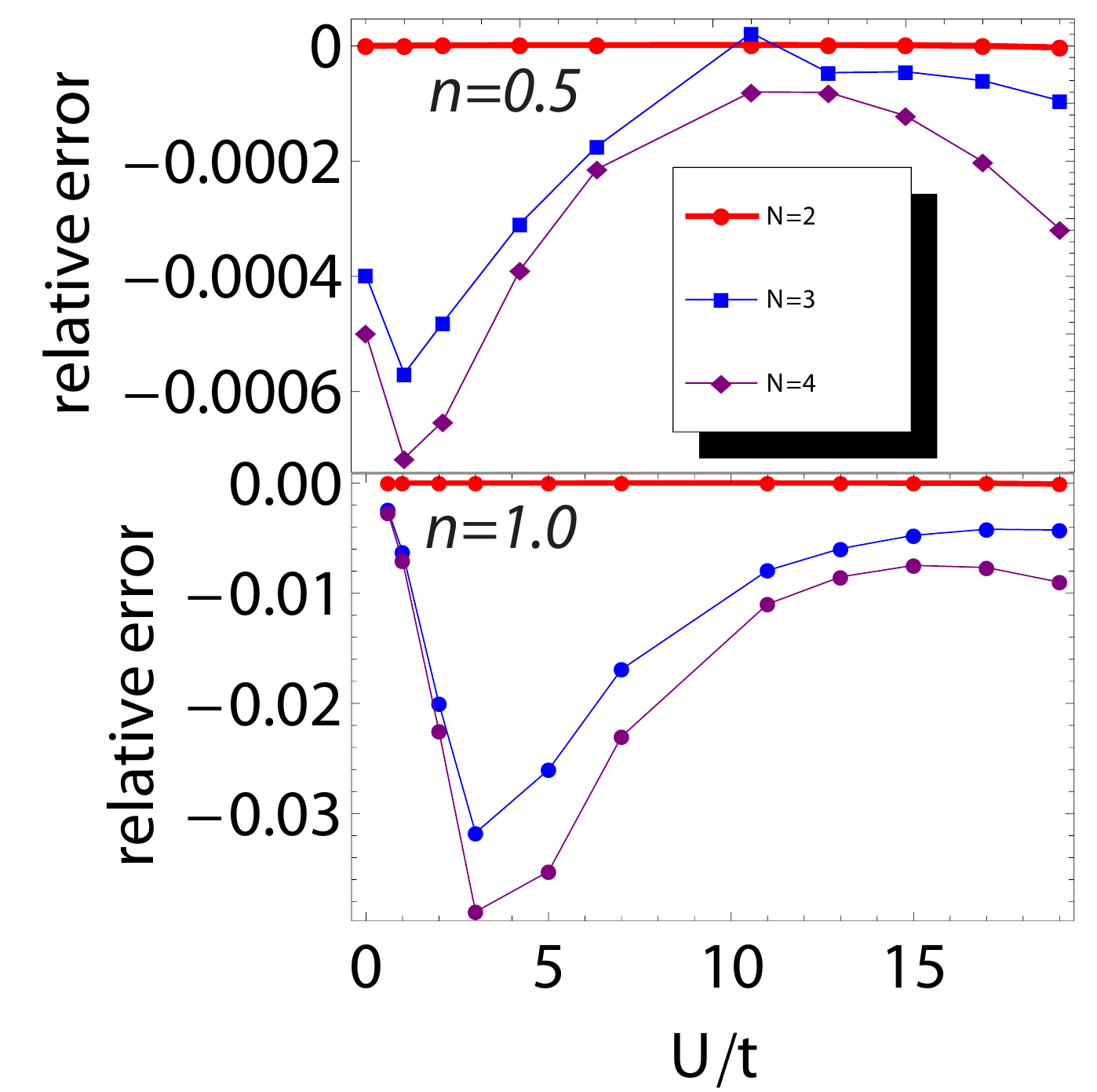}
\caption{(Color online) Relative error between DMRG and Bethe ansatz for $N = 2, \, 3, \, 4$ (from top to bottom) at densities $n=0.5$ ($1/(2N)$ filling, top) and at $n=1$ ($1/N$ filling, bottom).}
\label{fig:compareDMRGBA}
\end{figure}

Figure~\ref{fig:compareDMRGBA} presents the relative difference between the DMRG and the BA results for the ground state energies per site in the thermodynamic limit for $N \leq 4$.
We find that the relative error for $N=4$ is $\lesssim 4\%$ at $1/N$ filling, $n=1$, and $\lesssim 0.7\%$ at $1/(2N)$ filling $n=0.5$. 
For $N=2$, the relative error is of the order of $10^{-4}$ or smaller and is hence not visible on the scale of the plot.
This is expected since here the BA is exact, and the DMRG is known to be capable of obtaining the ground state energy for finite systems with a relative error of $10^{-6}$ or better \cite{Bedurftig:1998p457,dmrgbook}.
For $N>2$, the BA becomes exact for $U=0$ and in the limit $U\to\infty$, as explained above in Sec.~
\ref{sec:BAtreatment}. 
We hence expect the absolute errors
to be maximal for some intermediate value of $U$.
As can be seen in Fig.~\ref{fig:compareDMRGBA}, the maximal relative error is obtained at $U\approx 3t$.
The upturn at large-$U$ is discussed below.

The fraction of sites with three or more particles per site can be estimated in the non-interacting limit.
For densities $n\le2$, interactions suppress the number of triple and higher occupancies, so that the non-interacting limit yields an upper bound to the number of such configurations, and thus an upper  bound of the corresponding error.

\begin{figure}[t]
\includegraphics[width=0.5\textwidth]{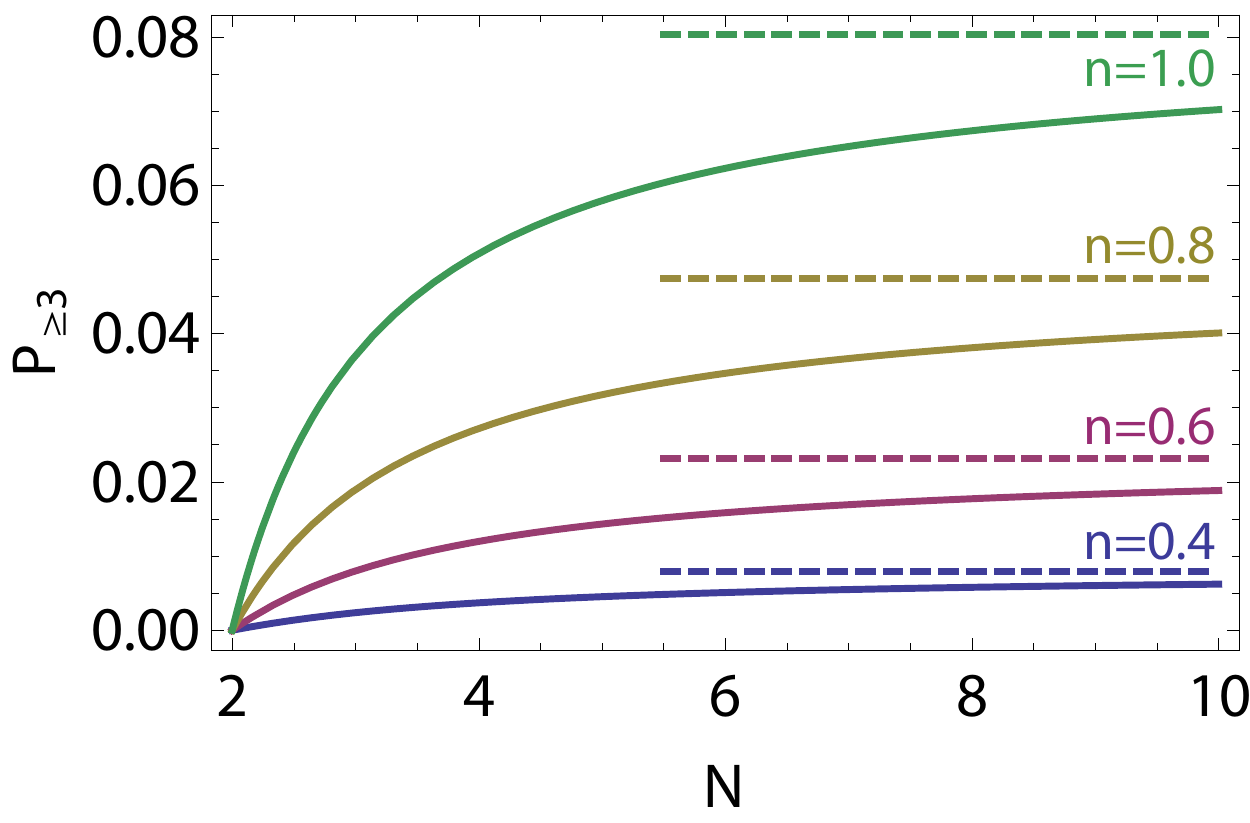}
\caption{(Color online) Triple or higher occupancy ($P_{\geq 3}$) versus $N$ for various densities  ($n=0.4,\ldots, 1.0$, bottom to top in steps of $0.2$).}
\label{fig:NItripocc}
\end{figure}

In the non-interacting limit, each of the $N$ flavors is independently occupied on a site with probability $n/N$, so that the probability of having $m$ particles per site is 
$
{}_mC_N 
 (n/N)^m (1-n/N)^{N-m}$, with ${}_mC_N = N!/[m! (N-m)!]$. 
Thus the probability of having three or more atoms on a  site is
\be
\hspace{-0.15in}P_{\geq3}(N) &=& 1-\sum_{m=0}^2 {}_mC_N (n/N)^m (1-n/N)^{N-m} \nonumber \\
    &&\hspace{-0.55in}{}= 1-\lp 1- n/N\rp^N\nonumber \\
    &&\hspace{-0.4in}{}\times\frac{ \lb n^2(N-2)(N-1)+2n(N-2)N + 2N^2 \rb}{2(N-n)^2}.
\ee
This function monotonically increases with $N$ and converges to its maximum as $N\rightarrow\infty$, given explicitly by
\be
P_{\geq 3}(N=\infty)      &=& 1-\frac{e^{-n}}{2}(n^2+2n+2).
\ee
Fig.~\ref{fig:NItripocc} plots the probability of triple and higher occupation $P_{\geq 3}(N)$ for $N=3$ and $N=\infty$ as a function of $N$ for various densities.
We see that at density $n=1$, the fraction of triple and higher occupancies for $N=3$ is $P_{\geq 3}(3)=0.037$ and for $N=\infty$ is $P_{\geq 3}(N=\infty)=0.080$.

Equation~\eqref{eq:3ptcl-BA} suggests that the relative error is on the order of the fraction of the triply or higher occupied sites times $U^2/(4t)$,  
approximating $f(\mathbf{k}) \sim 1$, which is a typical value, although  for $k_1 = k_2 = k_3 = \pi/2$ it diverges.
For $N=4$ at $n=1$, the quantity $P_{\geq 3} [U/(4t)]^2$ is .003 and .03 for $U=t$ and $U=35t$, in agreement with the results of Fig.~\ref{fig:compareDMRGBA} (bottom), about .005 and .04. 
Since this estimate grossly overestimates fluctuations in the strong coupling limit, 
we also estimate the fluctuations there for $n=1$.
Second order perturbation theory in $t/U$ gives fluctuations to three or more particles per site with amplitude $O\lp(t/U)^2\rp$, so Eq.~\eqref{eq:3ptcl-BA} suggests an error in the energy of the order $O(t)$. 
This absolute error becomes small in absolute terms as $t\rightarrow 0$, but the \textit{relative} error diverges since the exact energy scales as $t^2/U$.  Fig.~\ref{fig:compareDMRGBA} shows this upturn at large-$U$ of the relative error.

\section{Fidelity susceptibility as a probe for the Mott transition}
\label{sec:fidsusc}

In this section, we discuss the numerical results for the fidelity and fidelity susceptibility for systems with $N \leq 4$ at $1/N$ filling obtained via Eqs. (\ref{eq:fidelity}) and (\ref{eq:chi}).
We start by summarizing general considerations and former results for the Hubbard chain.

\subsection{Fidelity susceptibility at $U=0$: exact results}
\label{sec:chi_U0}

For small values of $U$, the fidelity susceptibility $\chi(U)$ [Eq.~(\ref{eq:chi})] can be obtained from standard perturbation theory.
For the SU$(N)$ Hubbard chain one obtains \cite{PhysRevLett.99.100603,PhysRevB.78.115410}
\begin{eqnarray}
\chi(U) & = &  \frac{1}{L} \sum_n \frac{|\langle n | V | 0 \rangle|^2}{E_n -E_0}  \\
V &=& \sum_{i, \alpha > \beta} n_{i, \alpha} n_{i, \beta} ,
\label{eq:perturbation1}
\end{eqnarray}
where $E_n$ and $| n \rangle$ are the eigenenergies and corresponding excited eigenstates of the SU$(N)$ Hubbard chain, and
$E_0$ and $| 0 \rangle$ are the energy and eigenstate of the ground state.

At the non-interacting point $U=0$, the ground state susceptibility can be computed exactly.
In momentum space,
we obtain the result in the thermodynamic
 limit \cite{PhysRevB.78.115410},
\begin{eqnarray}
&&\chi(U=0) =\frac{N(N-1)}{2 (2 \pi)^3} \nonumber \\
&& \times \int\limits_{-\pi}^{\pi} \int\limits_{-\pi}^{\pi} \int\limits_{-\pi}^{\pi} dk \, dk^\prime \, dq  \, \frac{n_k \,  (1- n_{k+q}) \, n_{k^\prime} \, (1-n_{k^\prime-q} )}
{ \left(\epsilon_{k+q}- \epsilon_{k}+ \epsilon_{k^\prime-q}- \epsilon_{k^\prime} \right)^2},
\label{eq:chiU0integral} 
\end{eqnarray}
with the single particle dispersion of non-interacting fermions $\epsilon_k = -2 t \cos (k)$ and $n_k = \Theta (\epsilon_F - \epsilon_k)$, with $\epsilon_F$ the Fermi energy.
The resulting numerical values of $\chi(U=0)$ for $N \leq 6$ are listed in Tab.~\ref{tab:chi}.
Note that for $N=2$ a {\it finite} value is obtained, indicating that there is no divergence of $\chi(U)$ at the metal-insulator transition. 
In addition, perturbation theory shows the derivative of $\chi(U)$ at $U=0$ is {\it negative} for $N>2$, demonstrating that a minimum is to be expected at some fine value of $U$.  
As we will see next, the critical point in the Hubbard chain is indeed characterized by such a minimum of $\chi(U)$.

\begin{table}[t]
\begin{tabular}{c|c}
$N$ & $\chi$\\
\hline
2 &  $1/(24 \pi^2) \approx 0.00422172$ \\
3 &  0.0109003 \\
4 &  0.0227492 \\
5 &  0.0416842 \\
6 &  0.0696197
\end{tabular}
\caption{Numerical values of the fidelity susceptibility $\chi$ at $U=0$ for $N = 2,\, \ldots, \, 6$
as obtained from Eq.~\eqref{eq:chiU0integral}. The $N=2$ case is
also shown in Ref.~\onlinecite{PhysRevE.76.022101}.}
\label{tab:chi}
\end{table}

\subsection{Scaling behavior and nature of $\chi(U=U_c)$ for SU$(N)$ Hubbard chains at  $1/N$ filling}
\label{sec:fsusc_scaling}

Reference~\onlinecite{PhysRevB.78.115410} analyzed the scaling behavior of $\chi$ for the SU$(2)$ Hubbard model at $1/N$ filling.
As suggested by the exact result at $U=0$ for $N=2$, $\chi(U)$ is found {\sl not} to diverge at the metal-insulator transition.
Applying Ref.~\onlinecite{PhysRevLett.99.095701}'s scaling argument to the SU$(N)$ Hubbard model,
we find that the singular part of the fidelity susceptibility goes to zero as one approaches the critical point:
the scaling exponent of the fidelity susceptibility near the critical point is given by $2\Delta - 2 z -1$
($\Delta = 2$ is the scaling dimension of the Hubbard interaction and $z=1$ is the dynamical exponent).
Therefore, the singular part of the fidelity susceptibility vanishes as one approaches the critical point.
Moreover, a large scale Quantum Monte Carlo calculation for the SU$(2)$ model in Ref.~\onlinecite{PhysRevE.76.022101}
finds that $\chi(U)$ has a local minimum at the critical point $U_c=0$. This indicates that
the regular part of the fidelity susceptibility behaves rather flat in the vicinity of $U_c = 0$.

The transition for $N>2$ at $1/N$  filling is believed to be of the same type as for $N=2$, but at a finite value of $U$ \cite{Assaraf:1999p728}.
Hence, it is natural to expect that also for $N>2$ the transition is identified by a local minimum of $\chi(U)$.
This is further corroborated by the fact that $K_{\rho}$ and the coupling of umklapp terms obey similar RG flow equations (of BKT type) for any value of $N$,
so that at $1/N$ filling it follows from Eq.~(\ref{eq:chi_Krho}) that $\chi$ should have similar behavior for all $N$ near the critical point.
Therefore, we expect a minimum of $\chi$ at $U_c$ for all values of $N$.
Note that the numerical computation of $\chi(U)$ is independent from the computation of $K_\rho$,
 which could show anomalous behavior at $U=0$ \cite{solyom_book,discussions}, and also independent
from the computation of the charge gap, from which it is difficult to obtain accurate values of $U_c$ due
to the exponential behavior at the BKT-type transition.

\subsection{Numerical results}
\label{sec:chi_numerics}

\begin{figure}[t]
\includegraphics[width=0.5\textwidth]{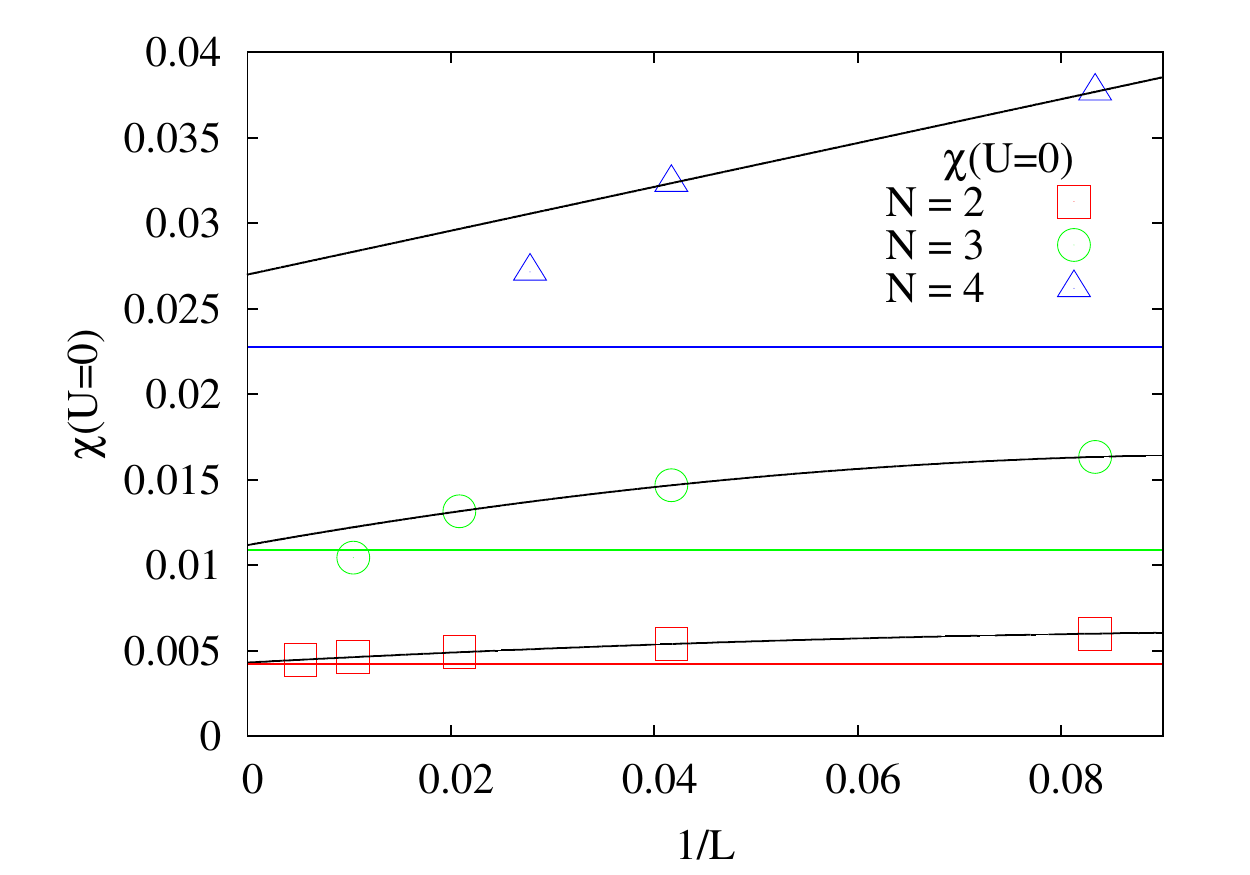}
\caption{(Color online) Finite size extrapolation of $\chi(U=0)$ for $N=2, \, 3, \, 4$. The horizontal lines show the exact values of Tab.~\ref{tab:chi}. For $N = 3$ and $N=4$ we have not taken into account the results for the largest system sizes ($L=96$ and $L=36$, respectively) due to a discarded weight of similar order of magnitude as $1-\mathcal{F}(U)$. The black lines show a quadratic fit for $N=2$ and $N=3$ and a linear fit for $N=4$.}
\label{fig:FS_U0}
\end{figure}

In order to identify $U_c$, we have computed the fidelity $\mathcal{F}_0(U) = |\langle \psi_0(U=0)|\psi_0(U)\rangle|$ for systems $L \leq 48$, 
and find no signature of a phase transition in the finite size data. 
We associate this to the BKT nature of the phase transition and expect that a discontinuity at $U_c$  should appear after extrapolating to the thermodynamic limit.
Due to the large numerical effort associated with such an analysis, we refrain from doing so and focus instead on the behavior of the fidelity susceptibility.

We start our discussion by estimating the accuracy of our numerical results by comparing to the exact results at $U=0$.
In Fig.~\ref{fig:FS_U0} we show our finite-size-scaling analysis for $\chi(U)$ and the comparison to the exact results of Tab.~\ref{tab:chi}.
We obtain $\chi(U=0)_{N=2} \approx 0.0043$ (exact value: $\chi \approx 0.00422$), $\chi(U=0)_{N=3} \approx 0.0112$ (exact value: $\chi \approx 0.01090$) and $\chi(U=0)_{N=4} \approx 0.02670$ (exact value: $\chi \approx 0.022749$).
Fig.~\ref{fig:FS_U0} also shows that for $N=3$ the results for $L=96$ lead to a bad extrapolation, and for $N=4$ the results for $L=36$ are not accurate enough for our considerations.
We therefore restrict the finite size scaling to $N=2$ and $N=3$, for which the relative error of the numerical results at $U=0$ is $< 3\%$, and discuss the qualitative behavior of the finite size data for $N=4$.

\begin{figure}[t]
\includegraphics[width=0.45\textwidth]{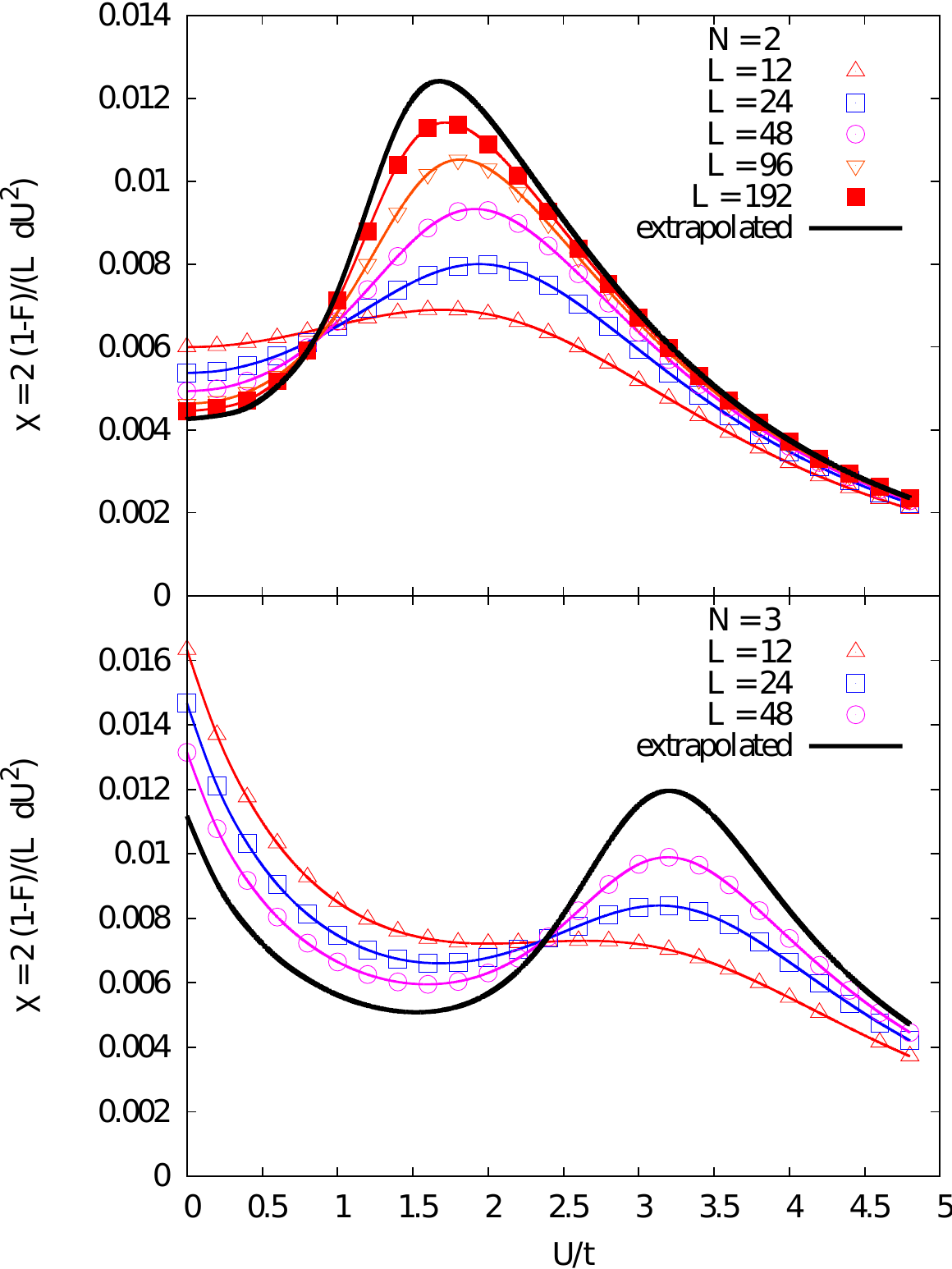}
\caption{(Color online) Fidelity susceptibility for different system sizes for $N=2$ and $N=3$. The black continuous line shows the value after extrapolating to the thermodynamic limit using the system sizes shown.
}
\label{fig:fidsusc}
\end{figure}

Figure~\ref{fig:fidsusc} shows $\chi(U)$ for systems up to $L=192$ ($N=2$) and $L=48$ ($N=3$), and the extrapolation to the thermodynamic limit.
The finite size results for $N=4$ show qualitatively similar behavior.
Interestingly, the results in Fig.~\ref{fig:fidsusc} show various similarities between $N=2$ and $N=3$.
In particular, in the thermodynamic limit, a minimum is obtained at $U_{\rm min} = 0$ for $N=2$ and $U_{\rm min} \approx 1.5t$ for $N=3$, followed by a maximum at $U_{\max} \approx 1.6t$ for $N=2$ and $U_{\rm max} \approx 3.2t$ for $N=3$.
Note that the numerical values of $\chi(U)$ at the minimum and at the maximum for $N=2$ and $N=3$ are very similar to each other.
This raises the question if these values might be universal for all values of $N$.

Finite size results for $N = 4$ for $L \leq 36$ indicate similar behavior, with $U_{\rm min} \approx 2t$ and $U_{\rm max} \approx 3.5t$, but the more difficult convergence inhibits obtaining the values of $\chi(U)$ at the minimum and the maximum in the thermodynamic limit.
Noteworthy is also the finding of a universal crossing point in Fig.~\ref{fig:fidsusc} between the minimum and the maximum of $\chi(U)$, whose explanation lies beyond the scope of the present paper.
We therefore leave these issues open for future research.  

The findings of Fig.~\ref{fig:fidsusc} are interesting.
As suggested by the analysis of Sec.~\ref{sec:fsusc_scaling}, they support values of $U_c = 0$ for $N=2$ (in agreement with the exact BA result), and $U_c \approx 1.5t$ and $U_c \approx 2t$ for $N = 3$ and $N=4$, respectively.
These values of $U_c$ can be contrasted to the findings of Ref.~\onlinecite{Assaraf:1999p728} in which the analysis of numerical QMC results for the charge gap indicate $U_c \approx 2.2t$ for $N=3$ and $U_c \approx 2.8t$ for $N=4$.
In agreement with the conclusions of Ref.~\onlinecite{Buchta:2007p713}, this indicates that the analysis of the charge gap tends to overestimate the value of $U_c$. 
However, our results for $\chi(U)$ for $N=2$ and $N=3$ indicate that the dependence of $U_{\rm min}$ on the system size for this quantity is rather weak, so that the extrapolation appears to be well controlled.
We believe therefore that the analysis of Ref.~\onlinecite{Buchta:2007p713} underestimates the values of $U_c$, and that indeed $U_c > t$ for $N>2$.
This is further corroborated by computing analytically $\chi(U)$ in the limit $U \to 0$ using perturbation theory.
For $N>2$, we find that $\chi(U)$ is finite and {\it decreases} with $U$, supporting a scenario in which the minimum is located at a finite value of $U$.
Note that the values of $U_{\rm min}$ are in rough agreement with the results of an approximate BA treatment of the metal-insulator transition \cite{schlottmann:metal-insulator_1992}, 
which finds $2.5t \lesssim U_c \lesssim 3.5t$ for $N=3,\ldots, \infty$, overestimating the value of $U_c$ for $N=3$.

We complement these considerations by another estimate of $\chi(U)$ as obtained from Eq.~(\ref{eq:chi_Krho}), which relates $\chi(U)$ to $K_\rho(U)$ for the Luttinger model. 
Numerically computing the derivative of $K_\rho(U)$ shows that $\chi(U)$ has indeed a minimum which is located at $U \approx 0$ for $N=2$, $U \approx 1.1t$ for $N=3$ and $U \approx 2.1t$ for $N=4$.
These values are in good agreement with the values of $U_{\rm min}$ obtained by directly computing $\chi(U)$, and we associate the discrepancy to the errors in the numerical computation of the derivative.

We finish this section by relating our findings to ongoing experiments.
Due to the smallness of the charge gap, it will be difficult to precisely locate $U_c$ in the experiments.
However, our analysis suggests that for all $N>2$ Luttinger-liquid (LL) behavior can be addressed by the experiments in the regime $U \lesssim t$, and that the transition to a Mott-insulator takes place at some value of $U \lesssim 4t$ for $N \to \infty$.
This suggests that for $N>2$ it should be possible to realize the LL by tuning $U$ to a small enough value, but not necessarily exactly to zero.
Also note that similar considerations imply that it should be possible to realize at finite temperatures $T$ spin-incoherent LL phases which we expect for $\hbar \, u_S / L \ll k_B \, T \ll \hbar \, u_C / L$, with $u_S$ and $u_C$ the velocity of the spin and charge excitations, respectively \cite{Review_spinincoherentLL}, which for small $U \ll t$ behave as $u_S \sim v_F - U/(2\pi)$ and $u_C \sim v_F + U\, (N-1)/(2\pi) $, where the Fermi velocity $v_F = 2t \sin(\pi/N)$.   

In the following section we will focus on the strong coupling limit of $1/N$ filled SU$(N)$ Hubbard chains, which maps to SU$(N)$ spin chains for $U$ large enough. 
From the considerations in this section, we expect this mapping to work for $U \gg 4t$.
In the following, we provide a more precise estimate by comparing energies and correlation functions.

\section{Heisenberg limit of SU$(N)$ Hubbard chains}
\label{sec:largeU}

\begin{figure}[b]
\includegraphics[width=0.48\textwidth]{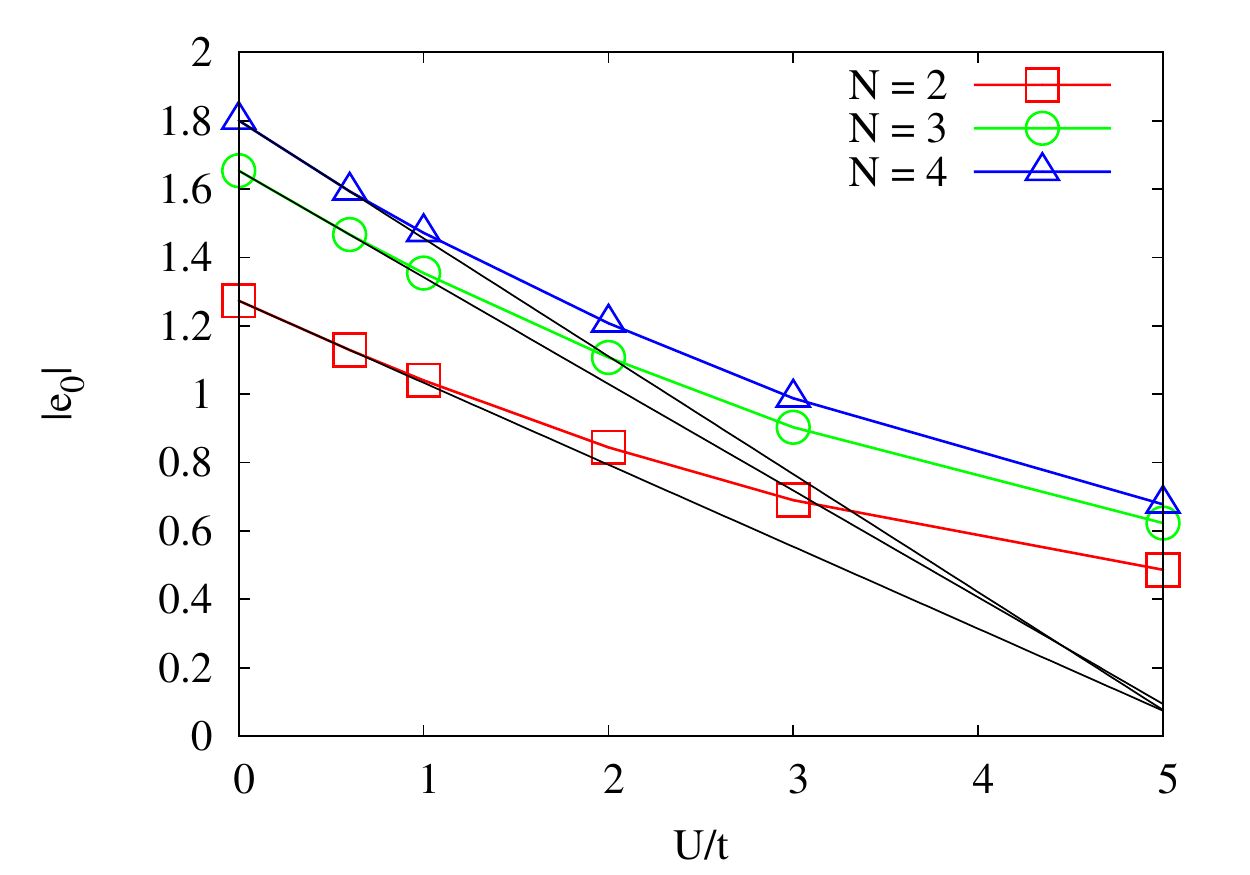}
\includegraphics[width=0.48\textwidth]{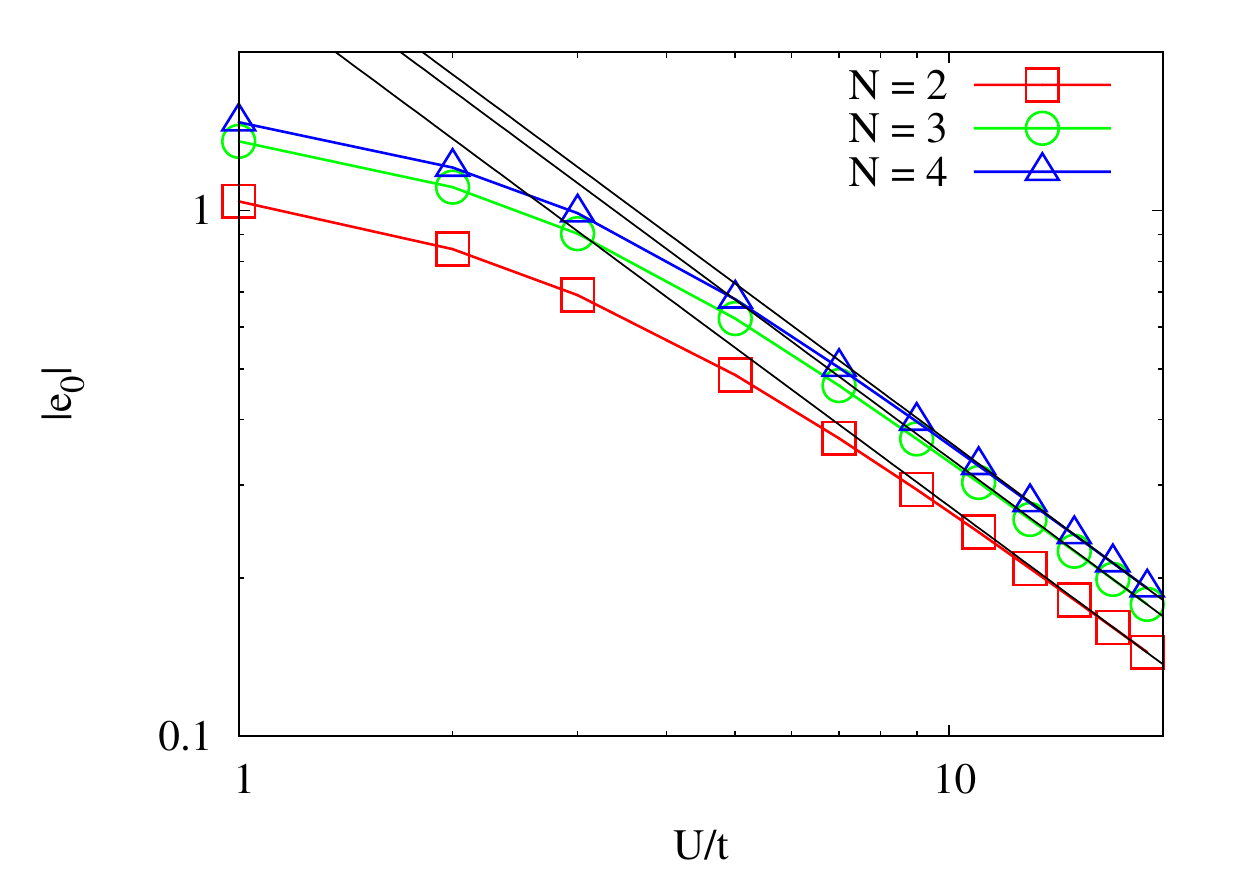}
\caption{(Color online) DMRG results for the absolute value of the ground state energy per site $e_0$ in the thermodynamic limit. Top: weak coupling regime. The black lines are linear fits to the energy using the first two data points.  Bottom: crossover region to the strong coupling regime (log-log scale).
The black lines are fits to a function $\sim t^2/U$ using the last two data points. 
}
\label{fig:DMRGlimits}
\end{figure}

\subsection{Energies}
\label{sec:HubbardHeisenbergEnergies}

\begin{figure}[t]
\includegraphics[width=0.45\textwidth]{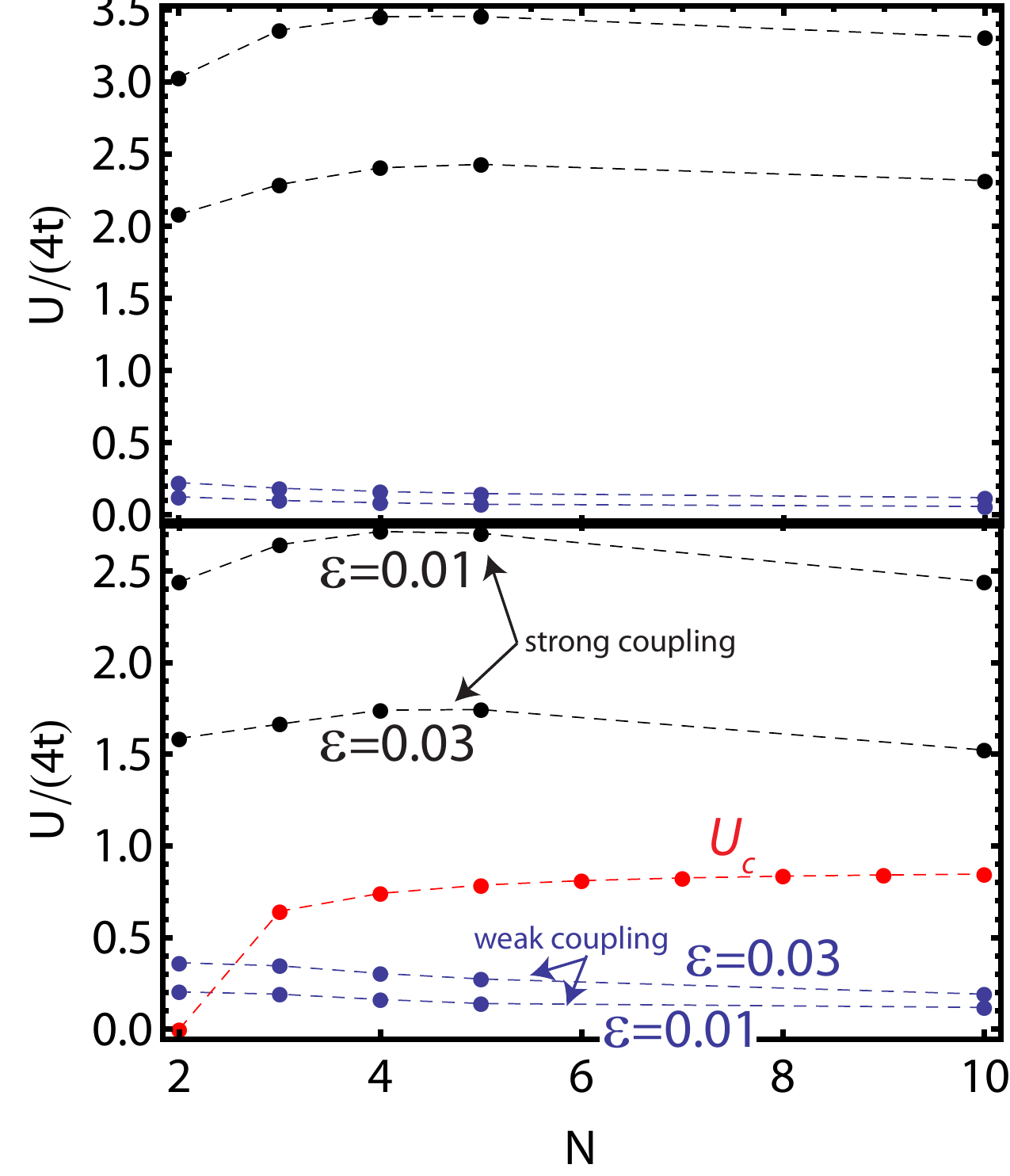}
\caption{(Color online) Black lines (top two curves in each panel) indicate where relative error of the ground state energy is within 1\% (lower black line) and 3\% (upper black line) of the $t^2/U$  asymptote. Blue lines (bottom two curves in each panel) indicate where the relative error of the ground state energy is within 1\% (upper blue line) and 3\% (lower blue line) of the weak-coupling perturbation theory. The red curve (extra middle line in bottom panel) indicates the critical $U_c$ for the Mott transition within the Bethe ansatz. Top: $1/(2N)$ filling ($n=0.5$), Bottom: $1/N$ filling ($n=1$).}
\label{fig:regions}
\end{figure}

It is \textit{a priori} unclear for which values of $U/t$ the SU$(N)$ Hubbard model behaves as a Heisenberg model at low energies, especially in the light of the probable differences in $U_c$ for $N=2$ and $N>2$ discussed in the previous section.
We address this by first comparing the DMRG energies of the SU$(N)$  Hubbard models to the expected $\sim t^2/U$ behavior in Fig.~\ref{fig:DMRGlimits} and find that, for the values of $N$  shown, the Heisenberg regime starts at $U_S \approx 11t$.
At this value, the difference of the DMRG results for $N=2$ to the expected $t^2/U$ behavior is $\epsilon \approx 5\times 10^{-3}$.
Note that $U_S$ shows a slight {\it decrease} upon increasing $N$ indicating that for all values of $N$ and  for $U \gtrsim 11t$ the system behaves as a Heisenberg chain.
This is further confirmed by BA, which shows that for $N \leq 10$ the energy follows the $t^2/U$ behavior in this regime, as shown in Fig.~\ref{fig:regions}.
Note, however, that the BA shows a slight increase of $U_S$ with $N$.
The behavior of the energies hence suggests that in 1D, for all values of $N$, SU$(N)$ Heisenberg  physics can be realized in the experiments with ultracold alkaline earth atoms for $U \geq 11t$.
We will further refine this in the next section, where we compare the numerical values of spin correlation functions of both models as a function of $U/t$.
Note that this numerical value of $U_S$ is in good agreement with the findings of Ref.~\onlinecite{Yang:2010p998} for a frustrated 2D system, in which effective spin models are found to describe the SU$(2)$  Hubbard model on the triangular lattice for $U\gtrsim 10t$.
We therefore expect that the SU$(N)$ Heisenberg model may quantitatively describe experiments with alkaline earth atoms in optical lattices for $U>U_S$ also in higher dimensions.

\subsection{Higher local occupancies}
\label{sec:Ntuples}

One specific property of the SU$(N)$ Hubbard systems is	that each site can be populated with up to $N$ particles. 
In the context of a possible realization of SU$(N)$ Heisenberg physics, it is therefore interesting to analyze the strong-coupling behavior of the higher local occupancies. 
These quantities are accessible in experiments and can provide valuable information. 
More specifically, measurements of the $P_m$, the number of particles on sites with occupancy $m$, will allow to obtain all the moments of the density as well as related quantities of interest  
such as the photoassociation rate \cite{lett:photoassociative_1995}, which is $\propto \langle n(n-1) \rangle$ (with $n = \sum_\alpha n^\alpha$), and the rate at which atoms are lost from the trap, dominated by 3-body losses and hence  $\propto \langle n(n-1)(n-2) \rangle$.   
\textit{In situ} single-site resolved measurements directly give the parity, $\langle (-1)^n \rangle$, which may also be obtained from the $P_m$ \cite{Bakr:2009p2641,Bakr:2010p1984,Sherson:2010p2701,Trotzky:2010p930,Weitenberg:2011p2530}.  
Most informatively, one can use RF spectroscopy to directly measure $P_m$ \cite{campbell_imaging_2006}, in particular by quenching the state of interest to a deep lattice.  
$P_m$ could also be measured using interaction blockade in an optical superlattice \cite{PhysRevLett.101.090404}. 
Moreover, it may be possible to extend \textit{in-situ} single-site resolution experiments capabilities to measure $P_m$ directly \cite{Simon:2011p2830}. 

\begin{figure}[t]
\includegraphics[width=0.5\textwidth]{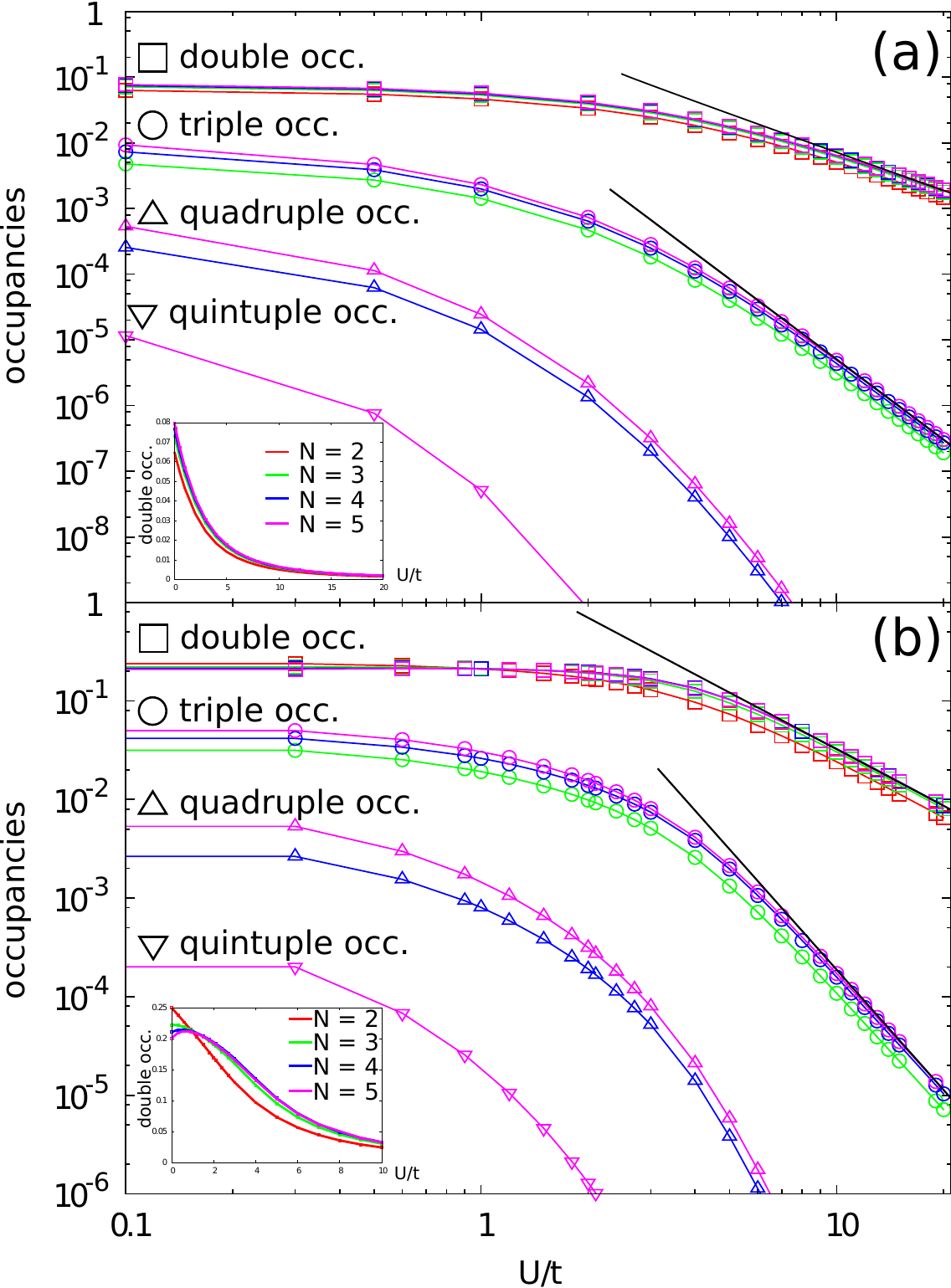}
\caption{(Color online) Site averaged double, triple, quadruple and quintuple occupancies as a function of $U/t$ for small systems at filling (a) $n=1/(2N)$, and (b) $n=1/N$. The system sizes in (a) are $L=24$ for $N=2,\, 3, \, 4$, and $L=20$ for $N=5$. The system sizes in (b) are $L=20$ for $N=2, \, 4$, $L=21$ for $N=3$, and $L=10$ for $N=5$. The black straight lines are guide to the eyes showing the power-law behavior. The insets show the double occupancy as a function of $U/t$ for the different values of $N$.}
\label{fig:Ntuples}
\end{figure}

In Fig.~\ref{fig:Ntuples} we show our DMRG results for systems with open boundary conditions and $L \leq 24$ for $N \leq 5$.  
More specifically, we present the average over all sites of the double occupancy $\langle D \rangle$,  the triple occupancy $\langle T \rangle$, the quadruple occupancy $\langle  Q_4   \rangle$ and of the quintuple occupancy $\langle Q_5 \rangle$, defined in Appendix~\ref{appendixB}. 
As can be seen, the results do not depend strongly on $N$. 
In particular in the limit of large $U/t$, we observe that the results for $N=4$ and $N=5$ are practically indistinguishable. 
In this limit, the quantities follow a power law $\sim (U/t)^{\eta}$, with exponents (obtained for $N=5$) $\eta_D \approx -1.9, \, \eta_T \approx -4.1 , \, \eta_{Q_4} \approx -7.9, \, \eta_{Q_5} \approx -12.0$ at filling $n=1/(2N)$ and $\eta_D \approx -2.0 , \, \eta_T \approx -4.0, \, \eta_{Q_4} \approx -7.9 , \, \eta_{Q_5} \approx -11.6 $ at filling $n=1/N$.  
Since the double occupancy is the largest quantity, we analyze it in more detail. 
The insets of Fig.~\ref{fig:Ntuples} show that the behavior at fillings $1/(2N)$ and $1/N$ differs at small values of $U/t$. 
While $\langle D \rangle(U)$ appears to decay monotonically for all values of $U/t$ at filling $1/(2N)$, at filling $1/N$ we identify for the small systems a rounding or a peak for $N>2$. 
This might be due to the metal-insulator transition. 
However, the positions of these maxima do not coincide with the minima of the fidelity susceptibility, so that we conclude that addressing the double occupancy in experiments with small systems is not sufficient to locate the phase transition. 
This can be understood since for the BKT transition all derivatives of the energy as a function of $U/t$ behave regularly. According to the Hellmann-Feynman theorem, the double occupancy is (to a good approximation for $N>2$) the first derivative of the energy with $U$, so that it is not expected to show singular behavior at the Mott-transition in these systems. 

However, all these quantities show a crossover to the aforementioned power-law behavior at values of $U \approx 10t$. 
Toghether with the behavior of the energy, this further supports that the minimal value of $U/t$ for  emulating Heisenberg physics to $\lesssim 1\%$ accuracy is approximately 10.  
We will now turn to the behavior of the correlation functions which further support this result.  

\subsection{Correlation functions}
\label{sec:correlresults}

\subsubsection{Bosonization results for the correlation functions and Luttinger parameters}
\label{sec:luttingerparameter}

According to Ref.~\onlinecite{Assaraf:1999p728}, and in agreement with our results for the fidelity
susceptibility discussed in Sec.~\ref{sec:fidsusc}, the system for $N>2$ is in a metallic (LL) phase at small,
but finite values of $U/t$.
Bosonization shows that at low energy the spin and charge degrees of freedom separate, and both sectors are described by the Luttinger liquids with corresponding Luttinger parameters $K_{\rho}$ (charge) and $K_\sigma$ (spin) \cite{giamarchi}.
Here, $K_\sigma = 1$ for any value of $U$. 
The leading order contributions to various correlation functions can be obtained from standard Abelian bosonization \cite{Assaraf:1999p728}.
We obtain for the density-density correlations [defined in Eq.~(\ref{eq:denscorrs})] 
\begin{equation}
\label{ncorr}
\langle N^{\rm total}(r) N^{\rm total}(0) \rangle  =  n^2 -\frac{N K_\rho}{2 \left( \pi r \right)^2}
+ A_1 \frac{\cos \left(2k_F r \right)}{r^{ 2K_\rho/N + 2 - 2/N}} ,
\end{equation}
where $k_F = n \pi/N$, and $n$ the density.
For the spin-spin correlations [Eqs.~(\ref{eq:spincorrs}) and~(\ref{eq:spincorrsheisenberg})] we obtain for $\alpha \neq \beta$
\begin{equation}
\langle  S_{\alpha}^{\beta} (r) S_{\beta}^{\alpha} (0)   \rangle =   -\frac{1}{2 \left(\pi r \right)^2} + B_1 \frac{\cos \left(2k_F r \right)}{r^{2 K_\rho/N + 2 - 2/N}},
\label{eq:SalphaBeta}
\end{equation}
and for $\alpha = \beta$,
\begin{equation}
\langle  S_{\alpha}^{\alpha} (r) S_{\alpha}^{\alpha} (0)    \rangle = \left( \frac{n}{N} \right)^2 -  \frac{ \left( K_{\rho}-1 \right)/N +1}{2 \left( \pi r \right)^2} + \frac{B'_1 \cos \left(2k_F r \right)}{r^{2 K_\rho/N + 2 - 2/N}},
\label{eq:SalphaAlpha}
\end{equation}
Note that due to the SU$(N)$ symmetry, Eq.~(\ref{eq:SalphaBeta}) is, up to a factor of 2, the same as $\langle S^z(r) S^z(0)\rangle$.
Also note that we neglect possible $4 k_F$ contributions and logarithmic corrections in the above expressions.
At $U = 0, \, K_\rho = 1$, and as the repulsive interaction $U$ is increased, the charge Luttinger parameter gradually decreases.
As discussed in Ref.~\onlinecite{Assaraf:1999p728}, at a sufficiently large value of $U = U_c$, the multiparticle umklapp scattering terms become relevant and a charge gap opens, leading to the metal-insulator transition of the  BKT type discussed in the previous sections.
In the Mott-insulating phase $U > U_c$, the spin correlations are then simply obtained from Eqs.~(\ref{eq:SalphaBeta}) and (\ref{eq:SalphaAlpha}) by setting $K_{\rho} =0$, and the density correlations decay exponentially.

Expression (\ref{ncorr}) can be used to obtain $K_\rho$ numerically.
In the limit $k\to0$ the charge structure factor behaves as
\begin{equation}
\mathcal N(k\to0) = \frac{N K_\rho}{2\pi} |k|;
\label{eq:Nksmallk}
\end{equation}
$K_\rho$ consequently can be determined by fitting the slope of the numerically obtained $\mathcal N(k)$ in the vicinity of $k=0$.

\subsubsection{DMRG results}

\begin{figure}[t]
\includegraphics[width=0.45\textwidth]{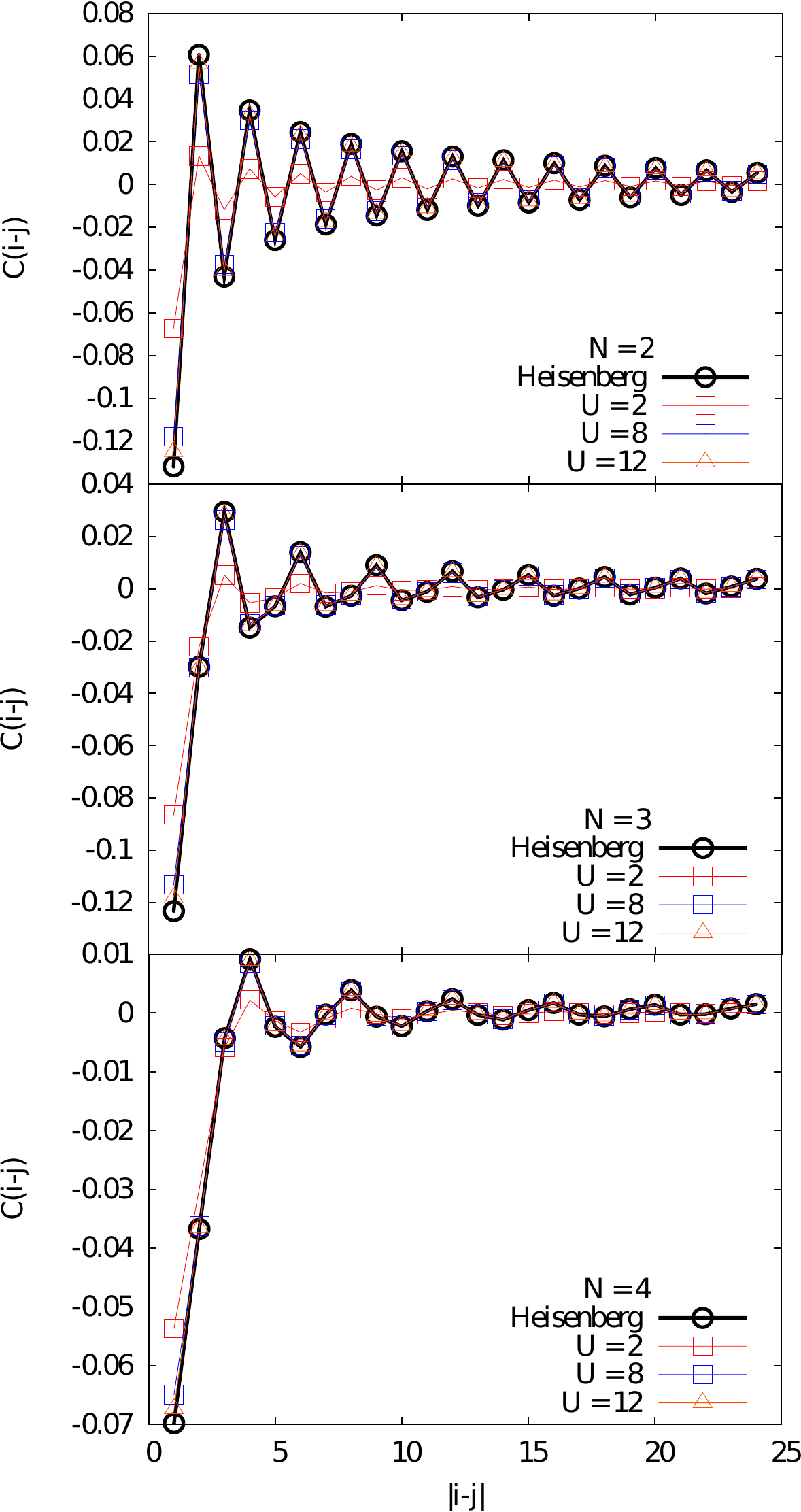}
\caption{(Color online) Comparison of spin correlation functions [Eqs.~(\ref{eq:spincorrs}) and (\ref{eq:spincorrsheisenberg})] of the $1/N$ filled SU$(N)$ Hubbard and SU$(N)$  Heisenberg chains at $U=2t,\, 8t, \, 12t$ for $N = 2, \, 3, \, 4$.}
\label{fig:correls}
\end{figure}

\begin{figure}
\includegraphics[width=0.485\textwidth]{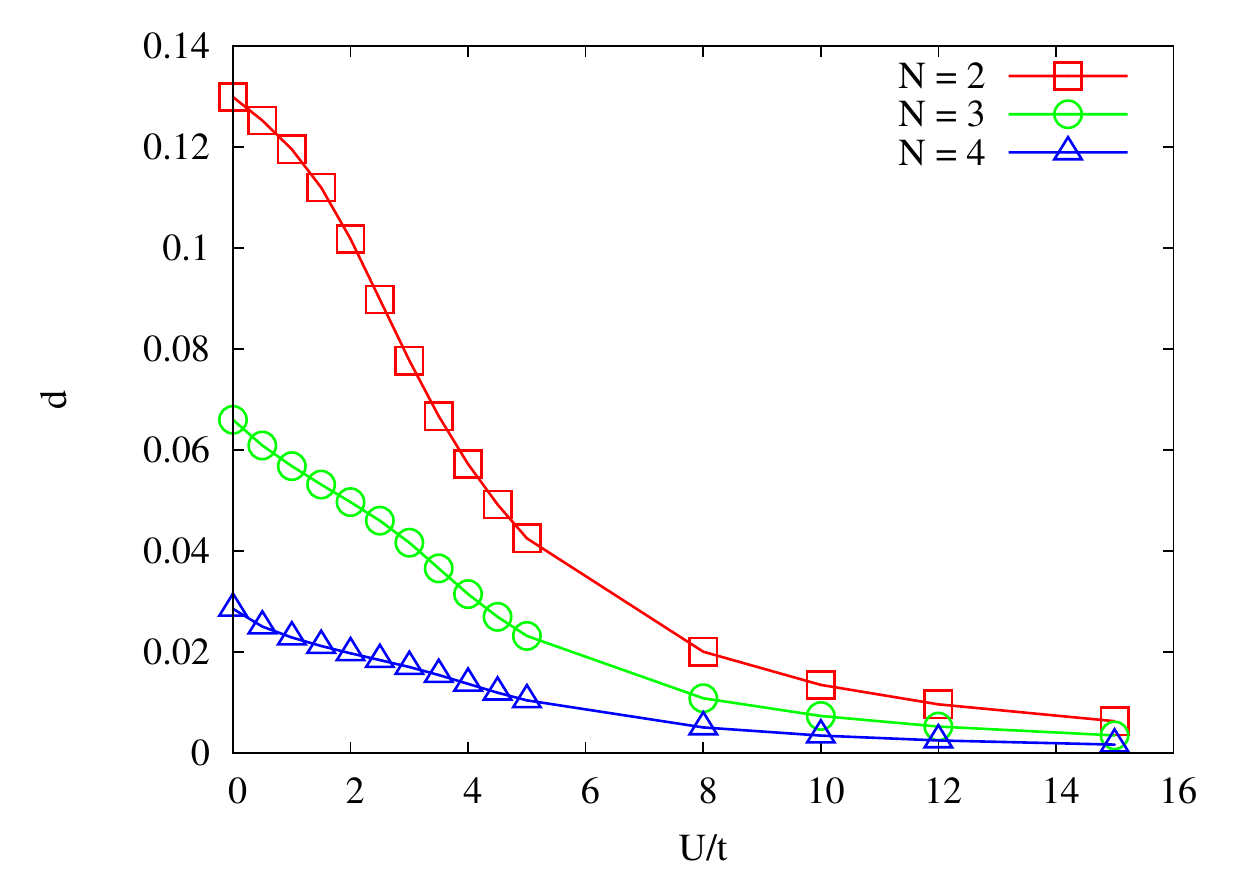}
\caption{(Color online) Distance $d$ [Eq.~(\ref{eq:distance})] between the spin correlation functions of the $1/N$ filled Hubbard and of the Heisenberg chains  [Eqs.~(\ref{eq:spincorrs}) and (\ref{eq:spincorrsheisenberg})] as a function of $U/t$ for $N \leq 4$.}
\label{fig:distancecorrels}
\end{figure}

In Fig.~\ref{fig:correls} we compare DMRG results for the spin correlation functions [Eq.~(\ref{eq:spincorrs})] of $N=2,3$, and $4$ SU$(N)$  Hubbard chains at unit filling for $U = 2t, \, 8t, \, 12t$  to the spin correlation functions of the corresponding Heisenberg chains [Eq.~(\ref{eq:spincorrsheisenberg})].
Already at $U=2t$ the Heisenberg model reproduces the qualitative features (algebraic decay and 2$k_F$ oscillations ) of the Hubbard model.
However, the difference in the actual values shows that this value of $U/t$ is outside the quantitative regime of validity of the Heisenberg model.
For $U=8t$ and $12t$, however, the agreement is quantitative for the three values of $N$ shown.
The largest difference is in the nearest-neighbor correlations.
Upon increasing $N$ the difference  decreases, corroborated
by computing the distance between the spin correlation function of the Hubbard systems [Eq.~(\ref{eq:spincorrs})] and of the Heisenberg systems [Eq.~(\ref{eq:spincorrsheisenberg})] which we define as
\begin{equation}
d = \sqrt{\sum_r \left[ S(r) - S^{\rm H}(r) \right]^2}.
\label{eq:distance}
\end{equation}
In Fig.~\ref{fig:distancecorrels}, we see that  this distance decreases with increasing $U/t$ and $N$. For $U > 12t$, we find $d < 0.01$ for all values of $N$.
Note that this criterion is matched for smaller values of $U/t$ when increasing $N$. \\

\begin{figure}
\includegraphics[width=0.485\textwidth]{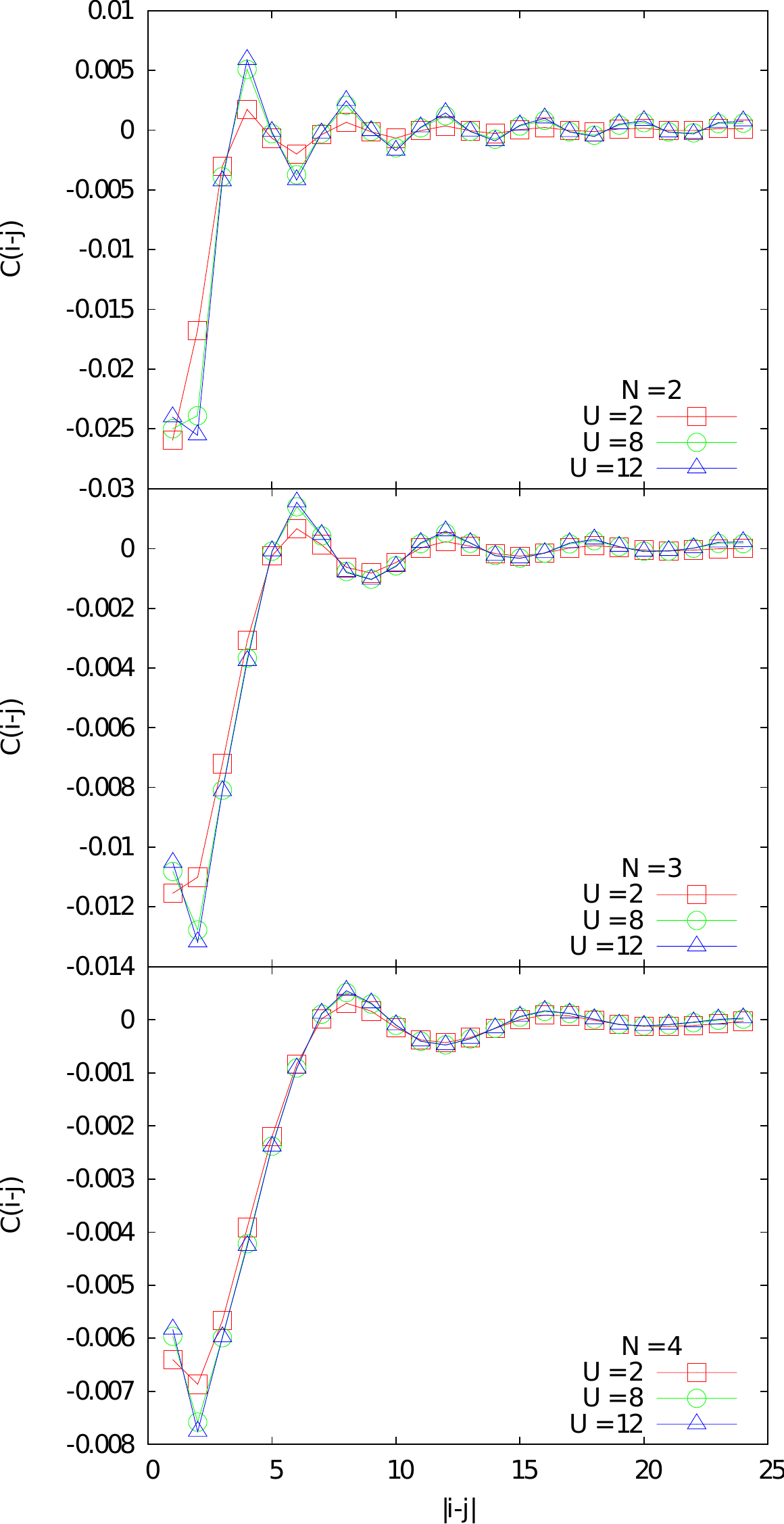}
\caption{(Color online) Spin correlation functions [Eq.~(\ref{eq:spincorrs})] of SU$(N)$ Hubbard chains at half filling $n=0.5$ for the same parameters as in Fig.~\ref{fig:correls}.}
\label{fig:half-filled-correlations}
\end{figure}

Figure~\ref{fig:half-filled-correlations} presents results for the spin correlation functions of the SU$(N)$ Hubbard chains at $1/2N$ filling.
The results show the same characteristics as at $1/N$ filling.
Interestingly, although this cannot be mapped to a Heisenberg model, the results for $U=8t$ and $U=12t$ are very similar to each other for the displayed values of $N$.
This suggests that the behavior in this region might be governed by SU$(N)$ t$-$J models.  These effective models capture the interplay of spin-exchange interactions expected for large values of $U/t$ with the electron itineracy.
In the SU$(2)$ case, it is known that this model possesses a rich phase diagram with superconducting phases \cite{PhysRevB.83.205113}.
In the SU$(N)$ case, the question arises if the phase diagrams of these models at $N=2$ and $N>2$ remain similar, as in the case of unit filled Hubbard chains, or if the enhanced symmetry might lead to unconventional phases, e.g., exotic singlet-superconductivity with singlets formed by $N$ particles.

 \subsection{Structure factors and Luttinger parameter}
 \label{sec:structurefactors}

\begin{figure*}[t]
\includegraphics[width=\textwidth]{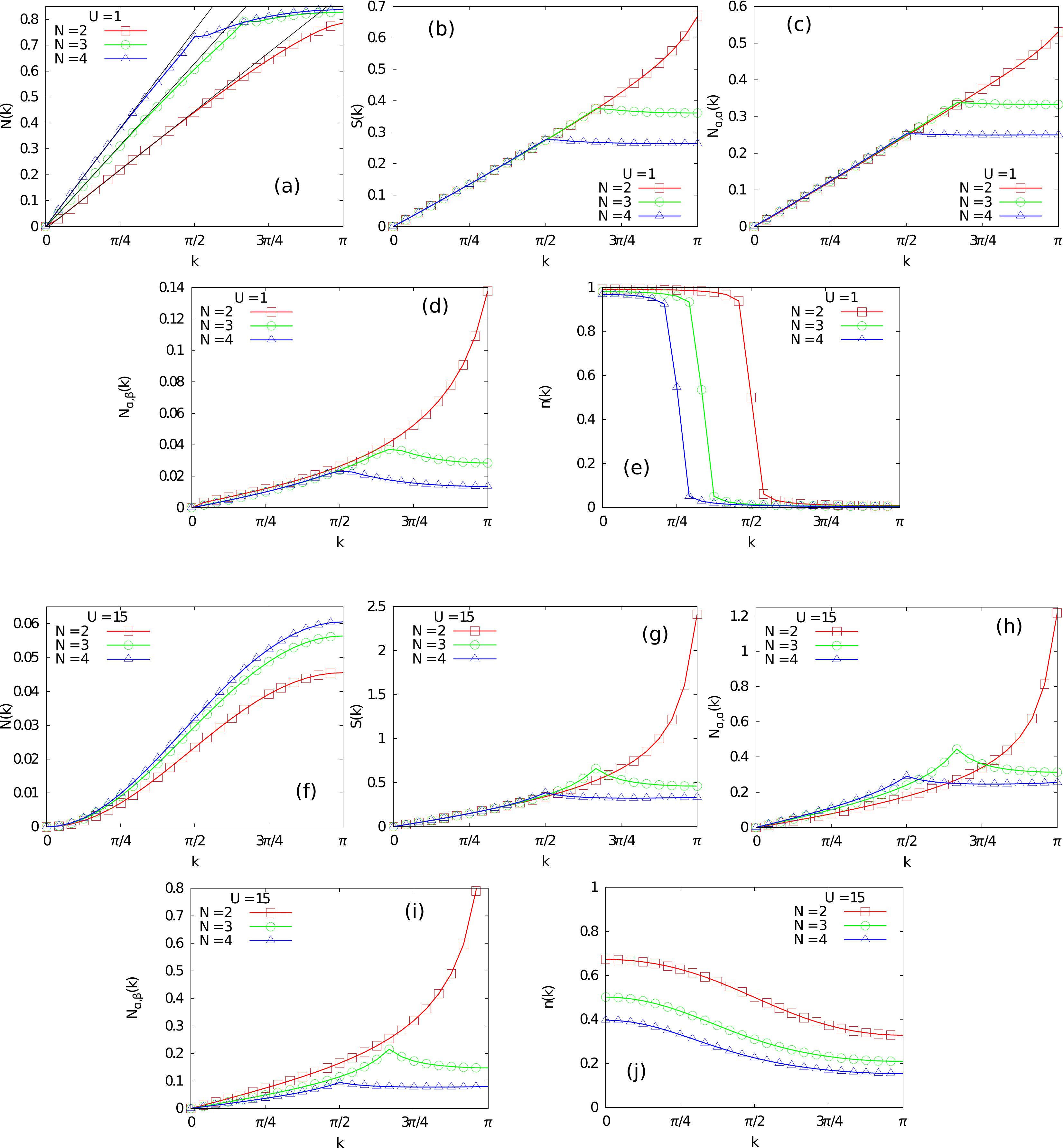}
\caption{(Color online) Structure factors of the various correlation functions Eqs.~(\ref{eq:spincorrs}) - (\ref{eq:OPDM}). (a) Charge structure factor [Eq.~(\ref{eq:denscorrs})] for $U=t$. (b) Spin structure factor [Eq.~(\ref{eq:spincorrs})] for $U=t$. (c) Structure factor $\mathcal{N}_{\alpha,\alpha}(k)$ [Eq.~(\ref{eq:alphaalphacorrs})] for $U=t$. (d) Structure factor $\mathcal{N}_{\alpha,\beta}(k)$ [Eq.~(\ref{eq:alphabetacorrs})] for $U=t$. (e) Momentum distribution function [Eq.~(\ref{eq:OPDM})] for $U = t$. (f)-(j): the same quantities, but for $U=15t$. }
\label{fig:sfactorsU1}
\end{figure*}

Figure~\ref{fig:sfactorsU1} shows our results for the various structure factors defined in Sec.~\ref{sec:correlfunctions} at $U=t$ and $U=15t$.
As expected from bosonization (see Sec.~\ref{sec:luttingerparameter}), all structure factors show a peak or a shoulder at $2 k_F$ originating from the oscillatory component of the correlation functions.
At $U=t$, the behavior of all structure factors at small $k$ is linear with $k$ up to $k \approx 2 k_F$ (only $\mathcal N_{\alpha,\beta}(k)$ shows a nonlinear behavior).
The momentum distribution function indicates the presence of a discontinuity; this is an artifact due to the small system sizes available, and for $N>2$ one would obtain a singularity in the derivative of $n(k)$ at $k_F$, according to LL theory.
For $N=2$, the results look similar due to the pronounced finite size effects caused by the exponentially slow opening of the charge gap.
This is also the reason why $\mathcal N(k)$ appears to be linear for $N=2$ despite the presence of a charge gap.
Below we will exploit the fact that for $U/t$ small enough, $\mathcal N(k)$ behaves linearly and obtain $K_\rho(U)$ from Eq.~(\ref{eq:Nksmallk}).

For $U=15t$, deep in the Mott-insulating phase, the linear behavior at small $k$ is less pronounced or absent due to the finite charge gap, which leads to an exponential decay of the correlation functions.
The most drastic changes are seen in $\mathcal N(k)$ and $n(k)$, which directly probe charge degrees of freedom.
In these quantities, the singularities at $2 k_F$ and $k_F$ disappear, as expected for a Mott insulator with a large charge gap.
Note that $\mathcal S(k)$ behaves linearly in the region $0 < k < \pi/4$, and is the same for all values of $N$ with a slope of $1/(2 \pi)$, in agreement with the bosonization result for the spin correlation function, Eq.~(\ref{eq:SalphaBeta}).

\begin{figure}[t]
\includegraphics[width=0.45\textwidth]{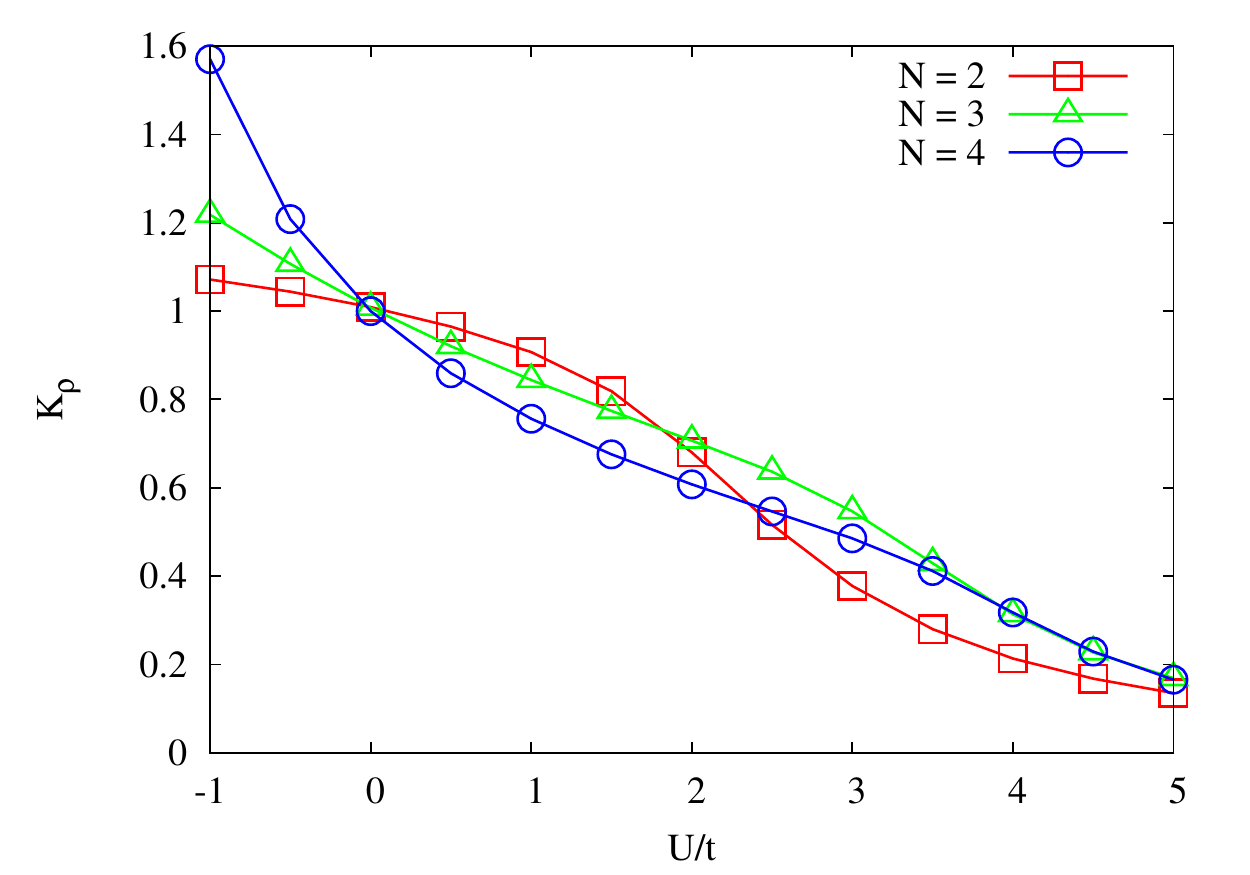}
\caption{(Color online) $K_\rho(U)$ as obtained for systems with $L=48$ sites via Eq.~(\ref{eq:Nksmallk}) for $N=2, \, 3,\, 4$.}
\label{fig:Krho}
\end{figure}

Fig.~\ref{fig:Krho} shows our results for $K_\rho(U)$ obtained from fitting the slope of $\mathcal N(k)$ at $k \approx 0$.
Note that for $U > U_c$, formally $K_\rho$ does not enter the correlation functions \cite{giamarchi}
and the structure factor cannot be used to determine $K_\rho$.
However, due to the exponentially slow opening of the gap, when the system size is much smaller than the correlation length
the structure factor for the finite systems appears to behave linearly so that we fit the slope also in these cases.
It appears that for $N=2$, there seems to be an inflection point
at $U=0$, which leads to a minimum of $\chi(U)$ computed from $K_\rho$ using Eq.~(\ref{eq:chi_Krho}).
For $N>2$, similar inflection points seem to appear at $U \approx 1.5t \, ($N=3$)$ and $U \approx 2t \, ($N=4$)$, i.e., close to the values of $U_{\rm min}$ at which the fidelity susceptibility $\chi(U)$ has its minimum.
Additional inflection points seem to appear at larger values of $U$ ($N=2$: $U \approx 2t$; $N=3$: $U \approx 4t$; $N=4$: $U \approx 4t$).
These are in rough agreement with the position of the maxima of $\chi(U)$ discussed in Sec.~\ref{sec:chi_numerics}. 

 \section{Summary and Conclusions}
 \label{sec:summary}
Using BA, bosonization, and DMRG, we have investigated SU$(N)$ Hubbard chains and identified the region of validity of both the strong- and weak-coupling perturbative regimes.
For $N>2$, where the BA is known to be an approximation, the values of the energies for $N=4$ agree with the ones obtained by DMRG with a relative error $< 4\%$.
We therefore use the BA to explore the behavior at large $N$ which is difficult to access with DMRG.
In addition, by computing the fidelity susceptibility, we have shed new light on the value of the critical interaction strength $U_c$ for the Mott transition at $1/N$ filling.
We identify a clear minimum in the fidelity susceptibility in the vicinity of the putative $U_c$ even for rather small system sizes.
Since the same behavior is obtained for $N=2$, for which $U_c=0$ is known exactly, we conclude that $U_c>t$ for all $N>2$.
For experiments with $N>2$, this signifies that it should be possible to observe Luttinger liquid behavior for $U<t$ even at $1/N$  filling.

For large $U/t$, we identify that SU$(N)$ Heisenberg models provide a very accurate ($\lesssim 1\%$ error) description of the system for $U> 12t$ for all values of $N$.
For these values of $U/t$, the absolute difference between the correlation functions of the Hubbard and the Heisenberg systems is $< 0.01$.
We expect therefore that in this regime the SU$(N)$ Heisenberg model will {\it quantitatively}  describe experiments with ultracold alkaline earth atoms on optical lattices.
We expect that also in higher dimensions this may be true for similar values of $U/t$.
Given that it is more favorable for the experiments to work with values of the spin-exchange interaction $J$ as large as possible, we therefore suggest to search for the proposed chiral spin liquid state \cite{Hermele:2009p1034} in experiments on square lattices for values of the interaction $10t \lesssim U \lesssim 15t$.

\section*{Acknowledgements}
We acknowledge useful discussions with Michael Hermele, Victor Gurarie, \"Ors Legeza, Jen\"o S\'olyom, Edina Szirmai and Simon F\"olling, as well as funding by NSF (PFC, PIF-0904017, and DMR-0955707), the AFOSR, and the ARO (DARPA-OLE). We also acknowledge CPU-time at ARSC. KRAH and GC would like to thank the Aspen Center for Physics, which is supported by NSF, for its hospitality during the writing of this paper. 

\begin{widetext}
\appendix
\section{Bethe ansatz equations}
\label{sec:appendixA}

Here we summarize the basic approximate Bethe ansatz equations developed in Refs.~\onlinecite{choy:some_1980,haldane:sun-bethe_1980}.

The rapidity distributions $\rho_c$ are associated with the charge degree of freedom, and the $\rho_j$'s for $j\in\{1,\ldots, s\}$ are associated with the rapidity distribution governing the difference in spin states $j-1$ and $j$ (where we interpret spin state ``0" as the charge).
These distributions  satisfy the coupled set of $N$ linear integral equations
\be
\rho_c(k) &=& \frac{1}{2\pi} + \frac{\cos(k)}{2\pi} \int_{-\Lambda_1}^{\Lambda_1} \!d\Lambda \, K_1(\sin k - \Lambda) \rho_1(\Lambda) \nonumber \\
\rho_1(\Lambda) &=&  \frac{1}{2\pi}\int_{-k_0}^{k_0}\!dk\,  K_1(\Lambda-\sin k)\rho_c(k) - \frac{1}{2\pi} \int_{-\Lambda_1}^{\Lambda_1} \! d\Lambda'\, K_2(\Lambda-\Lambda')\rho_1(\Lambda') \nonumber \\
    &&\hspace{0.4in}{}+\frac{1}{2\pi} \int_{-\Lambda_2}^{\Lambda_2} \! d\Lambda' \, K_1(\Lambda-\Lambda') \rho_2(\Lambda')  \nonumber \\
\rho_s(\Lambda) &=&  \frac{1}{2\pi} \int_{-\Lambda_{s-1}}^{\Lambda_{s-1}}\! d\Lambda'\, K_1(\Lambda-\Lambda') \rho_{s-1}(\Lambda') - \frac{1}{2\pi}\int_{-\Lambda_s}^{\Lambda_s}\! d\Lambda'\, K_2(\Lambda-\Lambda')\rho_s(\Lambda') \nonumber \\
    &&\hspace{0.4in}{}+\frac{1}{2\pi}\int_{-\Lambda_{s+1}}^{\Lambda_{s+1}}\!d\Lambda'\, K_1(\Lambda-\Lambda')\rho_{s+1}(\Lambda') \hspace{0.4in} \text{for } s=2,\ldots,N-2, \nonumber \\
\rho_{N-1}(\Lambda) &=&  \frac{1}{2\pi} \int_{-\Lambda_{N-2}}^{\Lambda_{N-2}}\! d\Lambda'\, K_1(\Lambda-\Lambda') \rho_{N-2}(\Lambda') - \frac{1}{2\pi}\int_{-\Lambda_{N-1}}^{\Lambda_{N-1}}\! d\Lambda'\, K_2(\Lambda-\Lambda')\rho_{N-1}(\Lambda')\label{eq:BA}
\ee
with
\be
K_q(x) &=& \frac{1}{2}\frac{qU}{(qU/4)^2+x^2}.
\ee

The parameters $k_0$ and $\Lambda_s$ for $s=1,\ldots,N-1$ are determined by the charge and spin densities $n_c$ and $n_j$ through
\be
n_c &=&  \int_{-k_0}^{k_0} \!dk\, \rho_c(k)  \nonumber \\
n_1 &=& \int_{-k_0}^{k_0} \!dk\, \rho_c(k) -\int_{-\Lambda_1}^{\Lambda_1}\!d\Lambda\, \rho_1(\Lambda)\nonumber \\
n_s &=& \int_{-\Lambda_{s-1}}^{\Lambda_{s-1}}\!d\Lambda\, \rho_{s-1}(\Lambda) \nonumber \\
    &&\hspace{0.05in}{}-\int_{-\Lambda_s}^{\Lambda_s}\!d\Lambda\, \rho_s(\Lambda) \hspace{0.25in} \text{for } s=2,\ldots,N-1 \label{eq:BA-densities-SUN-T0}
\ee
and the ground state energy per site is given by
\be
E_{\text{BA}} &=& -2t\int_{-k_0}^{k_0} \! dk\, \cos(k) \rho_c(k) \label{eq:BA-energy}
\ee

We numerically solve the integral equations, Gauss-Legendre discretizing the linear integral equations and solving the resulting linear equations~\cite{numericalrecipes}.
To apply the discretization procedure for finite intervals, we first transform the intervals $(-k_0,k_0)$ and $(-\Lambda_s,\Lambda_s)$ to $(-1,1)$.
A uniform rescaling of the coordinates is suitable for the $(-k_0,k_0)$ interval, but is unfavorable for the rest because $\Lambda_s$ tends to infinity for the population balanced gas, while the width of the function in the original units approaches a constant.
Thus, a uniform rescaling would require an unnecessarily large number of points as one would sample mostly where the integrand was zero. Instead, we rescale to new coordinates defined by
\be
\alpha(\Lambda) &=& \frac{\Lambda}{1+\Lambda}.
\ee
Defining the Jacobian
\be
j_{\Lambda}(u) &=& \Lambda \frac{1+\Lambda+\Lambda u^2}{(1+\Lambda-\Lambda u^2)^2},
\ee
the Bethe ansatz equations Eq.~\eqref{eq:BA} become
\begin{eqnarray}
\rho_c(q) &=& \frac{1}{2\pi}+\frac{\cos(k_0q)}{2\pi}\int_{-1}^1 \! du\, j_{\Lambda_1}(u) K_1\lp \sin (k_0q)-\frac{\alpha(\Lambda_1) u}{1-\alpha(\Lambda_1) u^2}  \rp \rho_1(u)\nonumber \\
\rho_1(u) &=& \frac{k_0}{2\pi} \int_{-1}^1\! dq\, K_1\lp \frac{\alpha(\Lambda_1) u}{1-\alpha(\Lambda_1) u^2} -\sin(k_0q)\rp \rho_c(q) \nonumber \\
    &&\hspace{0.3in}{}-\frac{1}{2\pi} \int_{-1}^1\! du'\, j_{\Lambda_1}(u')K_2\lp\frac{\alpha(\Lambda_1) u}{1-\alpha(\Lambda_1) u^2}-\frac{\alpha(\Lambda_1) u'}{1-\alpha(\Lambda_1) (u')^2}\rp \rho_1(u') \nonumber \\
           &&\hspace{0.3in}{}+\frac{1}{2\pi} \int\!du'\, j_{\Lambda_2}(u') K_1\lp\frac{\alpha(\Lambda_1) u}{1-\alpha(\Lambda_1) u^2}-\frac{\alpha(\Lambda_2) u'}{1-\alpha(\Lambda_2) (u')^2}\rp\rho_2(u')  \nonumber \\
\rho_s(u) &=& \frac{1}{2\pi}\int_{-1}^1\! du'\, j_{\Lambda_{s-1}} K_1\lp\frac{\alpha(\Lambda_s) u}{1-\alpha(\Lambda_s) u^2}-\frac{\alpha(\Lambda_{s-1}) u'}{1-\alpha(\Lambda_{s-1}) (u')^2}\rp \rho_{s-1}(u') \nonumber \\
    &&\hspace{0.3in}{}- \frac{1}{2\pi}\int_{-1}^1\! du'\, j_{\Lambda_{s}} K_2\lp\frac{\alpha(\Lambda_s) u}{1-\alpha(\Lambda_s) u^2}-\frac{\alpha(\Lambda_s) u'}{1-\alpha(\Lambda_s) (u')^2}\rp \rho_{s}(u') \nonumber \\
           &&\hspace{0.3in}{}+\frac{1}{2\pi}\int_{-1}^1\! du'\, j_{\Lambda_{s+1}} K_1\lp\frac{\alpha(\Lambda_s) u}{1-\alpha(\Lambda_s) u^2}-\frac{\alpha(\Lambda_{s+1}) u'}{1-\alpha(\Lambda_{s+1}) (u')^2}\rp \rho_{s+1}(u')  \hspace{0.4in} \text{for } s=2,\ldots,N-2,\nonumber \\
\rho_{N-1}(u) &=& \frac{1}{2\pi}\int_{-1}^1\! du'\, j_{\Lambda_{N-2}} K_1\lp\frac{\alpha(\Lambda_{N-1}) u}{1-\alpha(\Lambda_{N-1}) u^2}-\frac{\alpha(\Lambda_{N-2}) u'}{1-\alpha(\Lambda_{N-2}) (u')^2}\rp \rho_{N-2}(u') \nonumber \\
    &&\hspace{0.3in}{}- \frac{1}{2\pi}\int_{-1}^1\! du'\, j_{\Lambda_{N-1}} K_2\lp\frac{\alpha(\Lambda_{N-1}) u}{1-\alpha(\Lambda_{N-1}) u^2}-\frac{\alpha(\Lambda_{N-1}) u'}{1-\alpha(\Lambda_{N-1}) (u')^2}\rp \rho_{N-1}(u').
\end{eqnarray}

The relevant charge density, spin densities, and ground state energy are given by\\
\begin{eqnarray}
n_c &=& k_0 \int_{-1}^1\! dq\, \rho_c(q) \nonumber \\
n_1 &=& k_0 \int_{-1}^1 \!dq\, \rho_c(q) -\int_{-1}^1\!du \, j_{\Lambda_1}(u) \rho_1(u) \nonumber \\
n_s &=&  \int_{-1}^1 \! du\, j_{\Lambda_{s-1}}(u) \rho_{s-1}(u) \nonumber \\
    &&\hspace{0.1in}{}- \int_{-1}^1 \! du\, j_{\Lambda_{s}}(u) \rho_{s}(u) \hspace{0.25in} \text{for } s=2,\ldots,N-1, \nonumber \\
E_{\text{BA}} &=& -2t k_0 \int_{-1}^1 \! dq\, \rho_c(q).
\end{eqnarray}

Note that the BA results are obtained in the thermodynamic limit.
In Sec.~\ref{sec:compareDMRGBA}, we compare the results obtained by this procedure to the umerical results of the DMRG after extrapolating to the thermodynamic limit.

\section{Expressions for the higher local occupancies}
\label{appendixB}
In this appendix we define the expressions for the double occupancy $\langle D\rangle$, the triple occupancy $\langle T \rangle$, the quadruple occupancy $\langle Q_4 \rangle$, and the quintuple occupancy $\langle Q_5 \rangle$ discussed in Sec.~\ref{sec:Ntuples}.
We obtain:
\begin{eqnarray}
\langle D \rangle &=& \frac{1}{L} \sum\limits_{i=1}^{L} \left[ \sum_{\alpha, \alpha^\prime}   \langle n_i^\alpha n_i^{\alpha^\prime} \rangle 
- \Gamma^D_3 \sum_{\alpha, \alpha^\prime, \alpha^{\prime \prime}}  \langle n_i^\alpha n_i^{\alpha^\prime} n_i^{\alpha^{\prime \prime}} \rangle 
 - \Gamma^D_4   \sum_{\alpha, \alpha^\prime, \alpha^{\prime \prime}, \alpha^{\prime \prime \prime}}  \langle n_i^\alpha n_i^{\alpha^\prime} n_i^{\alpha^{\prime \prime}} n_i^{\alpha^{\prime \prime \prime}} \rangle  
-  \Gamma^D_5  \langle n_i^1 \, n_i^2 \, n_i^3 \, n_i^4 \, n_i^5 \rangle   \right] \label{eq:B1} \\
\langle T \rangle &=&  \frac{1}{L} \sum\limits_{i=1}^{L} \left[ \Gamma_3^{T} \sum_{\alpha, \alpha^\prime, \alpha^{\prime \prime}}   \langle n_i^\alpha n_i^{\alpha^\prime} n_i^{\alpha^{\prime \prime}} \rangle  - \Gamma^T_4   \sum_{\alpha, \alpha^\prime, \alpha^{\prime \prime}, \alpha^{\prime \prime \prime}}  \langle n_i^\alpha n_i^{\alpha^\prime} n_i^{\alpha^{\prime \prime}} n_i^{\alpha^{\prime \prime \prime}} \rangle  - \Gamma^T_5  \langle n_i^1 \, n_i^2 \, n_i^3 \, n_i^4 \, n_i^5 \rangle  \right] \\
\langle Q_4 \rangle &=&  \frac{1}{L} \sum\limits_{i=1}^{L} \left[ 
\Gamma_4^{Q_4} \sum_{\alpha, \alpha^\prime, \alpha^{\prime \prime}, \alpha^{\prime \prime \prime}}   \langle n_i^\alpha n_i^{\alpha^\prime} n_i^{\alpha^{\prime \prime}} n_i^{\alpha^{\prime \prime \prime}} \rangle 
-  \Gamma_5^{Q_4}  \langle n_i^1 \, n_i^2 \, n_i^3 \, n_i^4 \, n_i^5 \rangle \right] \\
\langle Q_5 \rangle &=&  \frac{\Gamma_5^{Q_5}}{L} \sum\limits_{i=1}^{L} \, \langle n_i^1 \, n_i^2 \, n_i^3 \, n_i^4 \, n_i^5 \rangle. \label{eq:B4}
\end{eqnarray}
The sums over the flavors $\alpha$ are over all possible permutations, and the numerical coefficients $\Gamma^o_p$ are listed in Tab.~\ref{tab:coefficients}.  
To obtain these coefficients, we choose them so that the N'th order polynomial of $n_i^\alpha$ reproduces the action of $P_m$ on a complete basis: $f({n^\alpha})\ket{m'} = P_m\ket{m'}=\delta_{mm'}$ for the $N+1$ values of  $m=0,\ldots,N$. 

\begin{table}[t]
\begin{tabular}{c|rrrrrrrrr}
            & $\Gamma_3^D$ & $\Gamma_4^D$ & $\Gamma_5^D$ & $\Gamma_3^T$ & $\Gamma_4^T$ & $\Gamma_5^T$ & $\Gamma_4^{Q_4}$ & $\Gamma_5^{Q_4}$ & $\Gamma_5^{Q_5}$ \\
\hline
$N=2$ & 0 & 0 & 0 & 0 & 0 &  0 & 0  & 0 & 0 \\
$N=3$ & 3 & 0  &  0 & 1 &  0 & 0  & 0 & 0 & 0\\
$N=4$ & 3 & -6  &  0 & 1 & 4 &  0 & 1  & 0 & 0 \\
$N=5$ & 3 & -6  &  10 & 1 &  4 & -10  & 1 & 5 & 1 
\end{tabular}
\caption{Numerical values of the coefficients in Eqs.~\eqref{eq:B1}-\eqref{eq:B4}.}
\label{tab:coefficients}
\end{table}

\end{widetext}



\end{document}